\documentclass[aps,prb,showpacs,twocolumn,amsmath,amssymb,floatfix,superscriptaddress]{revtex4-2}
\usepackage{amsmath}
\usepackage{amssymb}
\usepackage{graphicx}
\usepackage{bm}
\usepackage{color}
\usepackage{hyperref}
\usepackage{graphicx}
\usepackage{upgreek}
\usepackage{scalerel}
\usepackage{multirow}
\usepackage{footnote}

\usepackage{pgffor}
\usepackage{pdfpages}
\usepackage{mathdots}
\usepackage{float}
\usepackage{silence}
\usepackage{physics} 
\usepackage{lscape}   
\usepackage{color, colortbl}
\usepackage[table]{xcolor}
\usepackage{comment}  
\makeatletter
\AtBeginDocument{\let\LS@rot\@undefined}
\makeatother

\begin{document}
\title{$p$-wave superconductivity and Josephson current in $p$-wave unconventional magnet/$s$-wave superconductor hybrid systems}

\author{Yuri Fukaya}
\affiliation{Faculty of Environmental Life, Natural Science and Technology, Okayama University, 700-8530 Okayama, Japan}

\author{Keiji Yada}
\affiliation{Department of Applied Physics, Nagoya University, 464-8603 Nagoya, Japan}
 
\author{Yukio Tanaka}
\affiliation{Department of Applied Physics, Nagoya University, 464-8603 Nagoya, Japan}

\date{\today} 
\begin{abstract}
We study the surface density of states in $p$-wave unconventional magnet/spin-singlet $s$-wave superconductor hybrid systems ($p$-wave unconventional magnetic superconductors). 
Owing to the noncollinear spin structure in $p$-wave unconventional magnets, the spin-singlet $s$-wave pair potential behaves as the spin-triplet $p$-wave superconductivity. 
As a result, zero-energy flat bands can emerge at the edge.
Analyzing the pair amplitude at the edge, odd-frequency spin-triplet even-parity pairing is induced in the presence of zero-energy flat bands, while even-frequency spin-singlet even-parity remains.
We also demonstrate the Josephson current in superconducting junctions with $p$-wave unconventional magnet/spin-singlet $s$-wave superconductor hybrid systems.
By the cooperation of the spin-singlet $s$-wave pair potential and the $p$-wave unconventional magnetic order, the coupling of the spin-singlet even-parity pairings in junctions generates the first harmonics of the Josephson current.
In addition, the temperature dependence of the maximum Josephson current can be tuned by the chemical potential, which determines the generation of zero-energy flat bands.
Our results indicate that the $s+p$-wave-like superconducting state is generated in $p$-wave unconventional magnet/$s$-wave superconductor hybrid systems.
\end{abstract}
\maketitle



\section{Introduction}

The search for topological superconductivity~\cite{RevModPhys.83.1057,bernevig2013topological,schnyder2015,sato2016,sato2017topological,frolov2019quest,tanaka2012symmetry} plays an important role in the emergence of quantum computing for Majorana zero  modes~\cite{PhysRevB.82.134521,sarma2015majorana,aguado2020majorana,MarraJAP2022} characterized by zero-energy flat bands, which are topologically protected by winding numbers~\cite{Hu94,KT2000,YadaPRB2011, SatoPRB2011,BrydonPRB2011,schnyder2015,tanaka2024theory} and are relevant to odd-frequency superconductivity~\cite{tanaka2012symmetry}.
In particular, it is known that spin-triplet $p_{x}$-wave pairing leads to Majorana zero modes at the edge in one-dimensional superconductors (SCs) of the Kitaev chain model~\cite{kitaev2001}; however, spin-triplet $p$-wave SCs are not in general favorable in realistic materials.
For this perspective, a nanowire model, etc, with conventional $s$-wave SCs has been proposed~\cite{OregPRL2010,LutchynPRL2010,leijnse2012,LutchynPRL2011,KlinovajaPRL2013,san2013multiple}, and spin-triplet $p_{x}$-wave pairing is effectively induced to realize Majorana zero modes.
Indeed, the realization of the Majorana zero modes and zero-energy flat bands can be achieved by reducing the degree of freedom of an electron, which is caused by ferromagnetism, magnetic spin structure, like a spin helix, and so on~\cite{NakosaiPRL2012,KlinovajaPRL2013,EbisuPRB2015}, as well as by the proximity effect in superconducting junctions~\cite{IkegayaPRB2015,IkegayaJPCM2016,IkegayaPRB2017,IkegayaPRB2018,IkegayaPRB2020,IkegayaPRB2021_2,NagaePRB2024_2,zhu2025altermagnetic}.


It is known that the charge transport properties in superconducting junctions are strongly influenced by the edge states and spin structures of SCs~\cite{TK95,TKdJJ96,BarashPRL199,tanaka971,KT2000,Tanaka2004,TanakaGolubovPRL2007,tanaka2024theory,Kokkeler2025}.
In particular, these configurations determine the behavior of the current phase relation in Josephson junctions \cite{BarashPRL199,tanaka971}.
For instance, in conventional spin-singlet $s$-wave SC/spin-singlet $s$-wave SC Josephson junctions, without any edge states, the current phase relation is generally described by $\sum_{m}I_{m}\sin{(m\varphi)}$ with $I_{1}>0$ and the phase difference $\varphi$.
In the low-transparency limit studied by Ambegaokar and Baratoff,  $I_{1}\sin{\varphi}$ is satisfied, and the maximum Josephson current is saturated at low temperatures~\cite{Ambegaokar}.
While in spin-triplet $p$-wave SC/spin-singlet $s$-wave SC junctions, because spin-triplet pairing cannot be coupled to spin-singlet pairing within the first order of the tunneling process,
the $I_{1}$ component vanishes~\cite{PalsPRB1977,fenton1985,fenton1986,GeshkenbeinJETP1986,MillisPRB1988,yip1993,yamashiro1998,AsanoPRB2003}.
Besides, in $p_x$-wave SC/$p_x$-wave SC junctions, the Josephson current is enhanced by the resonance of zero-energy flat bands on both sides of the junctions~\cite{TanakaPRB1999,kwon2004fractional,AsanoPRL2006}.
However, in $p_x$-wave SC/$p_y$-wave SC junctions, the spin-triplet components can not be mixed within the first order owing to the different parity symmetries in the mirror plane along the interface~\cite{TanakaPRB1999,kwon2004fractional}.

\begin{figure*}[t!]
    \centering
     \includegraphics[width=17.5cm]{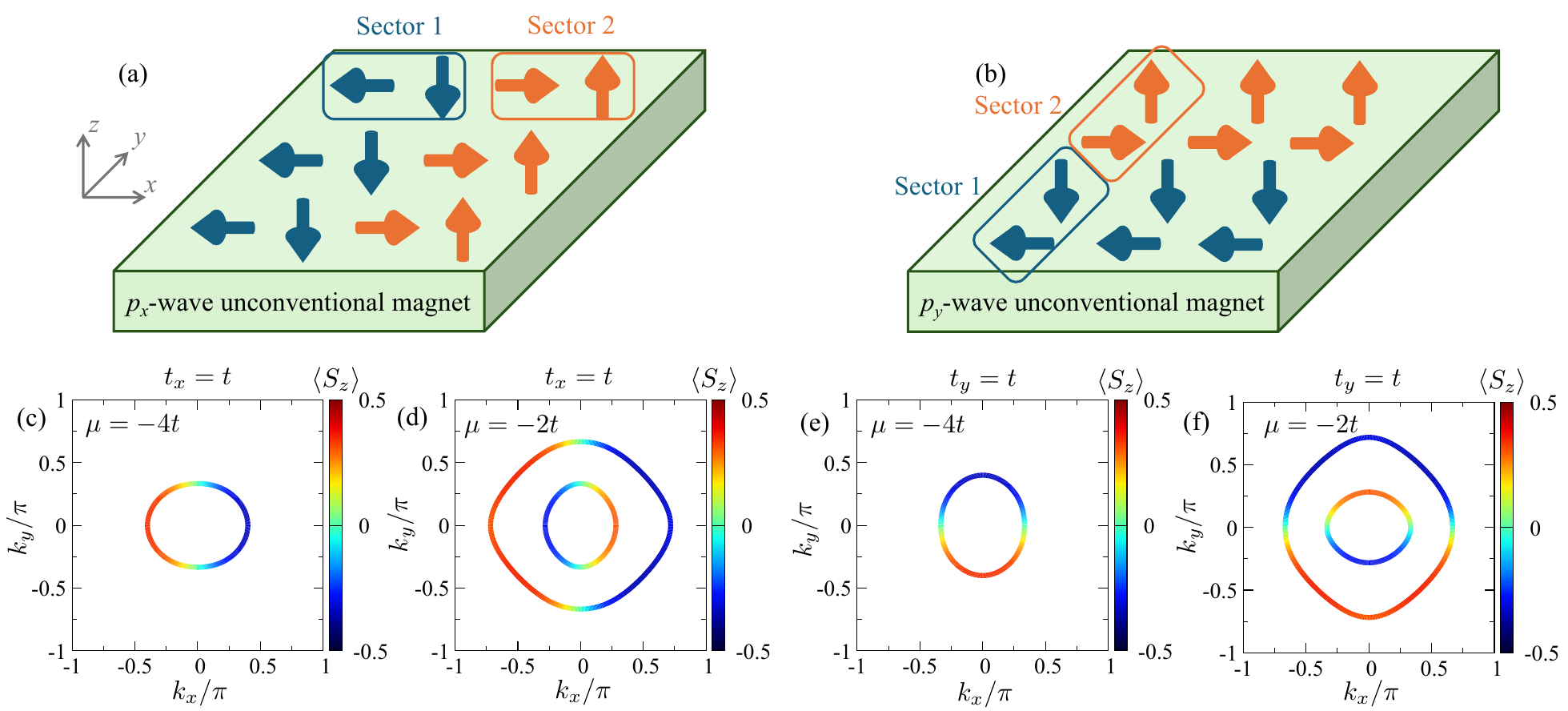}
    \caption{(a)(b) Schematic illustration of the two-dimensional $p$-wave unconventional magnet (PUM) at (a) $t_{x}\ne 0$ ($t_y=0$) and (b) $t_y\ne 0$ ($t_x=0$) in Eq.\ (\ref{PUM_eq}).
    Arrows stand for the noncollinear spin structure in PUM, and blue and red colors indicate the combination of the sectors, suggested in Ref.\,~\cite{BrekkePRL2024}.
    In the effective Hamiltonian, because two sectors (sectors 1 and 2) and spin degrees of freedom (up and down spins) contribute to the basis of the Hamiltonian~\cite{BrekkePRL2024}. 
    Both $J=t$ and $t_{x,y}$ terms lead to the noncollinear spin structures of $p$-wave unconventional magnetism.
    (c-f) Fermi surface at (c) $(\mu,t_{x},t_{y})=(-4t,t,0)$, (d) $(-2t,t,0)$, (e) $(-4t,0,t)$, and $(-2t,0,t)$.
    We set $J=t$ in panels (c)-(f).}
    \label{fig:1}
\end{figure*}%

Recently, a new class of magnetism has been discovered as ``altermagnetism''~\cite{noda2016momentum,NakaNatCommun2019,Hayami19,Ahn2019,NakaPRB2020,Yuanprb20,LiborSAv,landscape22,LiborPRX22,MazinPRX22,Bailing,Song2025,mazin2025notes,FukayaJPCM2025,liu2025review,LeePRL2025}, as well as odd-parity unconventional magnetism ~\cite{BrekkePRL2024,hellenes2024P,ChatterjeePRBL2024}.
In the former case, altermagnet/SC hybrid systems and superconducting junctions with altermagnetism exhibit Andreev reflection~\cite{Sun23,Papaj23,Niu_2024,nagae2024}, anomalous transport properties~\cite{Ouassou23,Beenakker23,Bo2024,fukaya2024,Cheng24,Wenjun25,SunPRB2025,Li2025nodelessAM,PalPRB2025}, nonreciprocal supercurrent~\cite{Banerjeediode24,chengdiode24,Chakraborty25,sharma2025diode,debnath2025diode,pal2025topological_diode,sharmadiode2026}, as well as topological superconductivity~\cite{cano23,Zhongbo23,CCLiu1,CCLiu2,LiPRB2024,mondal2024,hadjipaschalis2025Mzm,hodge2025platform,alam2025proximity,zhu2025altermagnetic,pal2025topological_diode,fu2025altermagnetism,mcardle2026topologicalsuper,sharma2026doublepeakmajorana,wan2026chiral}, etc~\cite{zhang2024,GiilPRB2024,ZyuzinPRB2024,SukhachovPRB2024,SukhachovPRB2024_2,SukhachovPRB2024_3,Maeda2025,ChourasiaPRB2025,leraand2025,heinsdorf2025,ChangPRB2025,monkman2025perscurrent,mukasa2025finite,vakili2026,madhusuthanan2026,zhao2026}.
In the latter case, $p$-wave unconventional magnetism is characterized by noncollinear spin structure~\cite{hellenes2024P,BrekkePRL2024,ChatterjeePRBL2024} and the resulting Fermi surface has a $p$-wave-like spin splitting.
As well as altermagnets, transport properties, and so on in $p$-wave unconventional magnet (PUM)/SC systems have been  studied~\cite{maeda24,fukaya2024,NagaePRB2025,fukaya2025PUM,SunZTPRB2025}.
Recently, some experiments have been successful in fabricating PUMs: NiI$_2$~\cite{song2025Nature} and Gd$_3$(Ru$_{(1-\delta)}$Rh$_\delta$)$_4$Al$_{12}$~\cite{yamada2025}.

It is known that ferromagnet/SC hybrid systems are the main topic of superconducting spintronics~\cite{linder2015NatPhys} and topological superconductivity~\cite{RevModPhys.83.1057,bernevig2013topological,schnyder2015,sato2016,sato2017topological,frolov2019quest,tanaka2012symmetry}, etc.
Apart from the ferromagnetism, because it is expected that unconventional magnetism does not break the superconductivity owing to the zero net magnetism, exotic superconducting phenomena can be led in hybrid systems with unconventional magnets and SCs.
In fact, altermagnetism in hybrid systems leads to higher-order topological superconductivity~\cite{LiPRB2024,Zhongbo23,CCLiu1,CCLiu2,mondal2024,hadjipaschalis2025Mzm}, subgap states~\cite{lu2025subgap}, and the crossed zero-energy flat bands~\cite{fukaya2025crossed}.
Likewise, in hybrid systems with a PUM, possible topological superconductivity and the edge states are proposed~\cite{ChatterjeePRBL2024,NagaePRB2025,SunZTPRB2025,khodas2026,pal2026,luo2026oddparity,kim2026,YiPRB2026,bera2026,DLliu2026}.
In Ref.~\cite{NagaePRB2025}, they adopted the effective PUM model proposed in Ref.\ \cite{hellenes2024P}, and zero-energy flat bands can be exhibited by tuning the chemical potential and $p$-wave magnetic order.
Indeed, studying hybrid systems composed of an unconventional magnet and SC is applicable for observing signatures of zero-energy flat bands in the tunneling spectroscopy experimentally.
Thus, it is timely to study not only charge transport behaviors but also pairing symmetry of hybrid systems with an unconventional magnet and SC. 

In this work, we demonstrate the emergence of $p$-wave superconductivity in PUM-spin-singlet $s$-wave SC hybrid systems: $p$-wave unconventional magnetic superconductors (PUM-SCs), by using the distinct PUM model~\cite{BrekkePRL2024}.
We show that zero-energy flat bands, which can be relevant to Majorana zero modes, can exhibit at the [100] edge owing to the noncollinear spin structure in the $zx$-plane, as also worked in Ref.\ \cite{NagaePRB2025}.
Then $p$-wave unconventional magnetic order leads to $p$-wave superconductivity associated with the spin-dependent hopping.
We analyze the pair amplitude at the [100] edge, and odd-frequency spin-triplet even-parity pairing can be realized in the presence of zero-energy flat bands, while even-frequency spin-singlet even-parity pairing also remains at the edge.
We also study the Josephson current in superconducting junctions with PUM-SCs along the $x$-direction.
Because the spin-singlet even-parity pair amplitude exhibits in PUM-SCs, the coupling of spin-singlet even-parity pairings on both sides of the junctions can occur and affect the current phase relation of the Josephson current.
Our results propose a more realistic implementation of zero-energy flat bands and the Josephson current, and broaden the possible methodology of the realization of spin-triplet $p$-wave superconductivity by the noncollinear spin structure of PUM.

The construction of this paper is as follows.
In Section II, we provide the model Hamiltonian and the Fermi surface in PUM.
We study the surface density of states (SDOS) at the edge in PUM-SCs in Section III and pair amplitude at the edge in Section IV.
We also study the Josephson effect in the high-transparency limit in Section V.
In Section VI, we show the temperature dependence of the maximum Josephson current in the low transparency limit.
Finally, we summarize and conclude our work in Section VII.

\section{Model Hamiltonian}

First, we present the model Hamiltonian in PUM.
The effective model of PUM has already been proposed in Refs.\ \cite{BrekkePRL2024,hellenes2024P}.
In this study, we deal with the tight-binding model suggested in Ref.\ \cite{BrekkePRL2024}:
\begin{align}
    \hat{H}(\bm{k})&=\varepsilon(\bm{k})\hat{s}_{0}\otimes\hat{\sigma}_{0}+[t_{x}\sin{k_x}+t_{y}\sin{k_y}]\hat{s}_{3}\otimes\hat{\sigma}_{0}\label{PUM_eq}\\
    &+J\hat{s}_{1}\otimes\hat{\sigma}_{3},\notag
\end{align}%
with the kinetic energy term $\varepsilon(\bm{k})=-\mu-2t\cos{k_x}-2t\cos{k_y}$, the intra-sector hopping $t$, the chemical potential $\mu$, local $sd$
interaction $J$, and the spin-dependent hopping $t_{x}$ and $t_{y}$.
Here, $\hat{s}_{0,1,2,3}$ and $\hat{\sigma}_{0,1,2,3}$ denote the Pauli matrices in spin and sector space, respectively.
In the Supplemental Material of Ref.~\cite{BrekkePRL2024}, the size of the original Hamiltonian in a minimal magnetic lattice is described as $8\times8$.
In Eq.\ (\ref{PUM_eq}), the $4\times 4$ Hamiltonian has been proposed to capture $p$-wave unconventional magnetism in the lattice model [Fig.~\ref{fig:1} (a) and Fig.~\ref{fig:1} (b).]~\cite{BrekkePRL2024}.
In this model, $t_{x}$ and $t_{y}$ terms generate the spin-helix structure~\cite{BerevigPRL2006,KohdaPRB2012,LiuPRL2014,IkegayaPRB2015,JacobsenPRB2015,YangPRB2017,AlidoustPRB2020,IkegayaPRB2021_2,AlidoustPRBL2021,LeePRB2021}, and the additional $J$ configuration leads to the noncollinear spin structure that characterizes $p$-wave unconventional magnetism shown in Fig.~\ref{fig:1} (a) and Fig.~\ref{fig:1} (b).
Then, $p_{x}$ ($p_{y}$)-wave magnetism is obtained by $t_{x}$ ($t_{y}$) with nonzero $J$~\cite{BrekkePRL2024}. 
Eq.\ (\ref{PUM_eq}) can be separated into two sectors:
\begin{align}
    \hat{H}^{\eta}(\bm{k})&=\varepsilon(\bm{k})\hat{s}_{0}+[t_{x}\sin{k_x}+t_{y}\sin{k_y}]\hat{s}_{3}+\eta J\hat{s}_{1},\label{PUM_eq_2}
\end{align}%
with the indices of the sector degrees of freedom $\eta=\pm1$, and the corresponding eigenvalues are given by
\begin{align}
    \bar{E}_{\pm}&=\varepsilon(\bm{k})\pm\sqrt{(t_{x}\sin{k_x}+t_{y}\sin{k_y})^{2}+J^{2}}.
\end{align}%
We show the Fermi surface of PUM with the expectation value of $S_{z}=(\hbar/2)\hat{s}_{z}$ at $\mu=-4t,-2t$, and $(t_{x},t_{y})=(t,0)$ and $(0,t)$ in Figs.~\ref{fig:1} (c-f).
These Fermi surfaces are doubly degenerate for each sector $\eta=\pm1$.
When we choose the parameters as $(t_{x},t_{y})=(t,0)$ shown in Fig.~\ref{fig:1} (c) ($\mu=-4t$) and Fig.~\ref{fig:1} (d) ($\mu=-2t$), we obtain the Fermi surface with $p_{x}$-wave unconventional magnetism, and then the number of Fermi surfaces are one and two, respectively.
Likewise, the Fermi surface with $p_y$-wave magnetism are demonstrated at $(t_{x},t_{y})=(0,t)$ [Fig.~\ref{fig:1} (e) for $\mu=-4t$ and Fig.~\ref{fig:1} (f) for $\mu=-2t$].
The obtained Fermi surface and the spin expectation values are the same as the result for $p_x$-wave UM cases with $\pi/2$-rotation. 
These demonstrations indicate that the direction of the noncollinear spin structure can be tuned by $t_{x}$ and $t_{y}$ terms.
It is noted that the expectation value of $S_{x}=(\hbar/2)\hat{s}_{x}$ ($\langle S_{x}\rangle$) in the normal state is canceled because its sign is opposite in each sector.

\begin{figure}[t!]
    \centering
    \includegraphics[width=8.5cm]{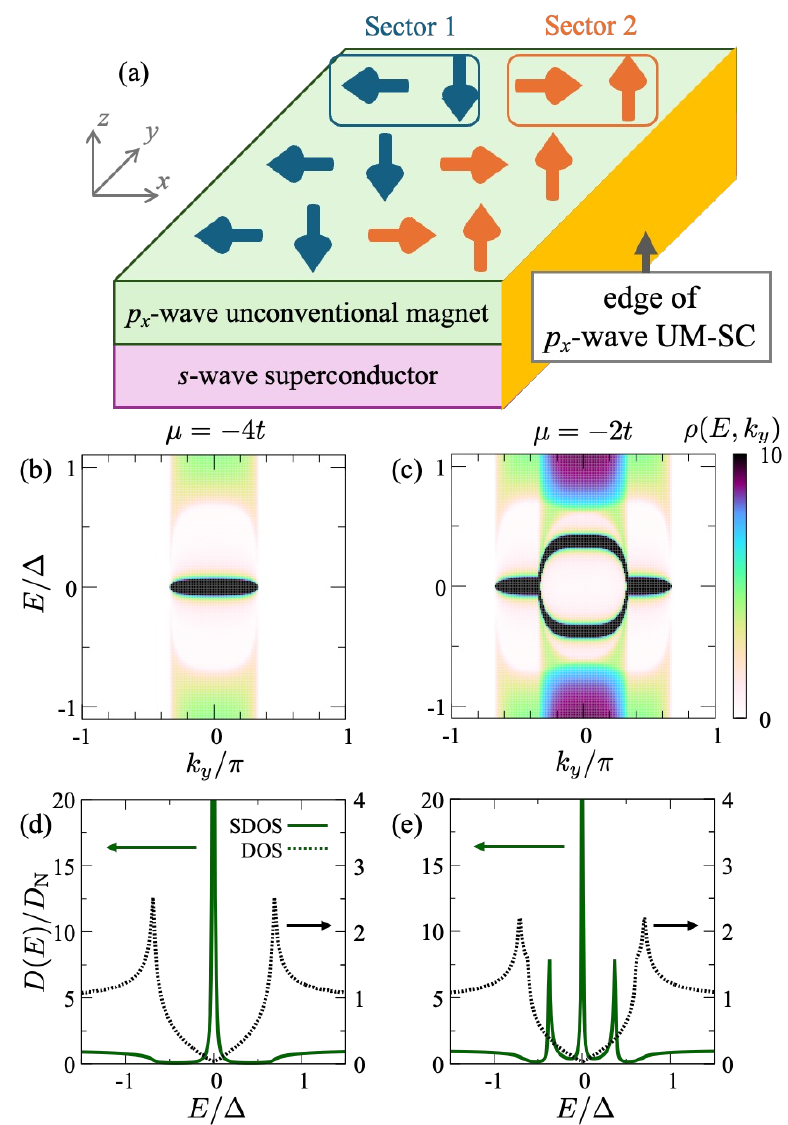}
    \caption{(a) Schematic illustration of PUM-$s$-wave SC hybrid systems ($p_x$-wave UM-SCs) at $t_{x}\ne0$ and $t_{y}=0$.
    We also show the two sectors in the effective Hamiltonian of Eq.\ (\ref{PUM_eq}).
    At the [100] edge, (b,c) momentum-resolved surface density of states (SDOS) $\rho(E,k_y)$ at (b) $\mu=-4t$ and (c) $\mu=-2t$.
    (d,e) SDOS (green solid) and bulk DOS (black dotted lines) normalized by $D_\mathrm{N}$ with the normal state SDOS and bulk DOS at zero energy for (d) $\mu=-4t$ and (e) $\mu=-2t$.
    Parameters: $(t_{x},t_{y})=(t,0)$, $J=t$, $\Delta=0.01t$, and $\delta=0.01\Delta$.}
    \label{fig:2}
\end{figure}

\begin{figure}[t!]
    \centering
    \includegraphics[width=8.5cm]{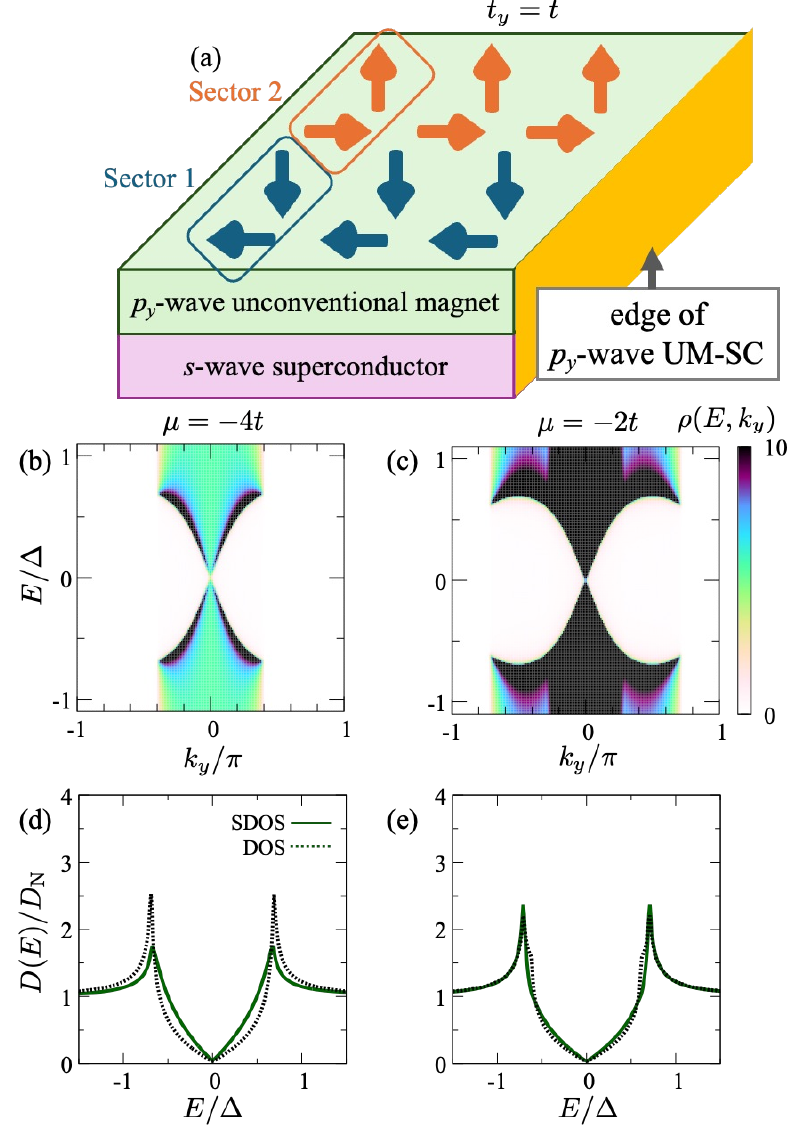}
    \caption{(a) Schematic illustration of PUM-$s$-wave SC hybrid systems ($p_y$-wave UM-SCs) at $t_{x}=0$ and $t_{y}\ne0$.
    We also show the two sectors in the effective Hamiltonian of Eq.\ (\ref{PUM_eq}).
    At the [100] edge, (b,c) momentum-resolved SDOS $\rho(E,k_y)$ at (b) $\mu=-4t$ and (c) $\mu=-2t$.
    (d,e) SDOS (green solid) and bulk DOS (black dotted lines) normalized by $D_\mathrm{N}$ with the normal state SDOS and bulk at zero energy for (d) $\mu=-4t$ and (e) $\mu=-2t$.
    Parameters: $(t_{x},t_{y})=(0,t)$, $J=t$, $\Delta=0.01t$, and $\delta=0.01\Delta$.}
    \label{fig:3}
\end{figure}

Here, we introduce the Bogoliubov-de Gennes Hamiltonian for PUM/$s$-wave SC hybrid systems (PUM-SCs) shown in Fig.~\ref{fig:2} (a) and Fig.~\ref{fig:3} (a) in the bulk~\cite{SukhachovPRB2024_3}:
\begin{align}
    \hat{H}_\mathrm{BdG}(\bm{k})=
    \begin{pmatrix}
        \hat{H}(\bm{k}) & \hat{\Delta}\\
        \hat{\Delta}^{\dagger} & -\hat{H}^{*}(-\bm{k})
    \end{pmatrix}.
    \label{BdG_eq}
\end{align}
We assume the conventional intra-sector spin-singlet $s$-wave pairing:
\begin{align}
    \hat{\Delta}=\Delta[\hat{s}_{0}\otimes\hat{\sigma}_{0}]i\hat{s}_{2}
\end{align}%
where $\Delta$ is the magnitude of the pair potential.
As topological superconductivity has been proposed in the nanowire model~\cite{OregPRL2010,LutchynPRL2010}, magnetic chain~\cite{KlinovajaPRL2013,EbisuPRB2015}, etc~\cite{NakosaiPRL2012}, spin-singlet $s$-wave pairing has been adopted for the proximity.
Based on these previous works and Ref.\ \cite{BlackSchafferPRB2008}, we choose the spin-singlet $s$-wave pair potential in PUM-SC.


\section{Surface density of states at the [100] edge}

In this section, we discuss the $p$-wave superconductivity and the emergence of the edge states in PUM-SCs.
For this purpose, we calculate the surface density of states (SDOS) by using the retarded Green's function $\hat{G}^\mathrm{R}(k_y,z)$ for  $z=E+i\delta$:
\begin{align}
    D(E)&=\int^{\pi}_{-\pi} \rho(E,k_y)dk_y,\\ 
    \rho(E,k_y)&=-\frac{1}{\pi}\mathrm{Im}\mathrm{Tr'}\tilde{G}^\mathrm{R}(k_y,z),
\end{align}%
with the energy $E$ and the infinitesimal value $\delta=0.01\Delta$.
We adopt the recursive Green's function method to obtain the semi-infinite Green's function $\tilde{G}^\mathrm{R}(k_y,z)$ in Refs.\,\cite{Umerski97,TakagiPRB2020}.

We focus on the SDOS at the [100] edge shown in Fig.~\ref{fig:2} (a) ($p_x$-wave UM-SC) and Fig.~\ref{fig:3} (a) ($p_y$-wave UM-SC).
In Fig.~\ref{fig:2} (b), Fig.~\ref{fig:2} (c), Fig.~\ref{fig:3} (b), and Fig.~\ref{fig:3} (c), we plot the momentum-resolved SDOS $\rho(E,k_y)$ at $(t_{x},t_{y})=(t,0)$ ($p_x$-wave UM-SC) [Fig.~\ref{fig:2} (b) and Fig.~\ref{fig:2} (c)], and $(t_{x},t_{y})=(0,t)$ ($p_y$-wave UM-SC) [Fig.~\ref{fig:3} (b) and Fig.~\ref{fig:3} (c)].
The chemical potential is chosen as $\mu=-4t$ [Fig.~\ref{fig:2} (b) and Fig.~\ref{fig:2} (c)], and $\mu=-2t$ [Fig.~\ref{fig:3} (b) and Fig.~\ref{fig:3} (c)].
Because noncollinear spin structure exhibits in the $zx$-plane for both $t_{x}$ and $J$ shown in Fig.~\ref{fig:2} (a), at $\mu=-4t$, both nodal points and zero-energy flat bands appear [Fig.~\ref{fig:3} (b)]~\cite{TanumaPRB2002_Sep,SenuptaPRB2001}.
While in Fig.~\ref{fig:3} (a),  noncollinear spin structure emerges in the $yz$-plane for both $t_{y}$ and $J$. 
Then the noncollinear spin structure aligns along the [100] edge, and as a result, we only obtain the nodal structure [Fig.~\ref{fig:3} (b)].
At $\mu=-2t$, for the $p_{x}$-wave UM-SC [$(t_{x},t_{y})=(t,0)$], zero-energy flat bands partly appear by connecting two nodal points shown in Fig.~\ref{fig:2} (c)~\cite{YadaPRB2011}, while for the $p_{y}$-wave UM-SC [$(t_{x},t_{y})=(0,t)$], only nodal structure appears [Fig.~\ref{fig:3} (c)].
We demonstrate the bulk energy gap structure (quasiparticle energy dispersions), and we confirm that the energy gap is closing in PUM-SCs and the nodal structure emerges for $J>\Delta$, in Appendix A. 
Indeed, these SDOS structures are effectively similar to the spin-triplet $p_x$ [Fig.~\ref{fig:2} (b)] and $p_y$-wave superconductivity [Fig.~\ref{fig:3} (b)].
It indicates that the $p$-wave superconductivity can be realized by PUM, even though we assume the spin-singlet $s$-wave pairing, and it is characterized by the spin-dependent hopping $t_{x}$ and $t_{y}$.
As clarified in Ref.\ \cite{NagaePRB2025}, we also obtain the zero-energy flat bands by using the distinct PUM model~\cite{BrekkePRL2024}.
We confirm the realization of zero-energy flat bands independent of the chemical potential $\mu$.
We note that these zero-energy flat bands are topologically protected by winding numbers~\cite{NagaePRB2025}.
These nodal structures protected by a topological invariant in bulk SCs host the zero-energy flat bands unless the energy gap closes by changing the parameters~\cite{SatoPRB2011}.
Thus, in our study, zero-energy flat bands caused by spin-triplet $p$-wave superconductivity can be realized within the parameter range.

After the integral for $k_y$, we obtain the SDOS and bulk DOS $D(E)$ normalized by the normal state SDOS and bulk at zero energy $D_\mathrm{N}$ at Fig.\,\ref{fig:2} (d) and Fig.\,\ref{fig:2} (e) $(t_{x},t_{y})=(t,0)$ ($p_x$-wave UM-SC), and Fig.\,\ref{fig:3} (d) and Fig.\,\ref{fig:3} (e) $(t_{x},t_{y})=(0,t)$ ($p_y$-wave UM-SC).
In the bulk, even though we only assume the spin-singlet $s$-wave pair potential, we obtain the V-shaped DOS for each case [black dotted lines in Figs.~\ref{fig:2} (d)(e) and Figs.~\ref{fig:3} (d)(e)].
This line shape originates from the nodal structure in the bulk, see also in Appendix A.
For the $p_{x}$-wave UM-SC [$(t_{x},t_{y})=(t,0)$], at $\mu=-4t$, in the presence of the zero-energy flat bands shown in Fig.~\ref{fig:2}(b), the zero-energy peak of SDOS is obtained [Fig.~\ref{fig:2}(d)]~\cite{TanumaPRB2002_Sep,SenuptaPRB2001}.
At $\mu=-2t$, three peaks of SDOS appear at $E=0$ and $E\sim\pm0.3\Delta$, as shown in Fig.~\ref{fig:2} (e).
We note that the peak at $E\sim\pm0.3\Delta$ originates from the flat bands of the SDOS at $E\sim\pm0.3\Delta$ in Fig.~\ref{fig:2} (d).
On the other hand, for the $p_{y}$-wave UM-SC [$(t_{x},t_{y})=(0,t)$], because zero-energy flat bands do not exhibit at $\mu=-4t,-2t$ [Fig.~\ref{fig:3}(b) and Fig.~\ref{fig:3}(c)], the V-shaped SDOS appears [Fig.~\ref{fig:3}(d) and Fig.~\ref{fig:3}(e)].
We note that this line shape originates from the nodal structure at $k_y=0$ [Fig.~\ref{fig:3}(b) and Fig.~\ref{fig:3}(c)].
Hence, zero-energy flat bands can be realized in PUM-SCs as the $p$-wave superconductivity.
These zero-energy flat bands can also be interpreted as topologically protected zero modes~\cite{PhysRevB.82.134521,sarma2015majorana,aguado2020majorana,MarraJAP2022}.

\section{Pair amplitude configurations at the [100] edge}

Since it is known that odd-frequency pairings emerge in zero-energy flat bands~\cite{TanakaPRL2007_1,Tanaka_Tanuma_PRB2007,tanaka2012symmetry,TamuraPRB2019_odd}, we here analyze the anomalous electron-hole Green’s functions, that is, the pair amplitude, at the [100] edge of PUM-SCs.
Pair amplitude $\hat{F}^{\sigma,\sigma'}_{j_x,j'_x}(k_y,i\omega_{n})$ with the spin $\sigma,\sigma'=\uparrow,\downarrow$, the site indices along the $x$-direction $(j_{x},j'_{x})$, and the fermionic Matsubara frequency $\omega_{n}=(2n+1)\pi k_\mathrm{B}T$ for the temperature $T$ at the [100] edge is obtained by the semi-infinite Green's function ($j_x=j'_x=0$) with
\begin{align}
    \tilde{G}(k_y,i\omega_{n})=
    \begin{pmatrix}
        \hat{G}(k_y,i\omega_{n}) & \hat{F}(k_y,i\omega_{n}) \\
        \bar{F}(k_y,i\omega_{n}) & \bar{F}(k_y,i\omega_{n})
    \end{pmatrix},
\end{align}
and the anomalous Green's part:
\begin{align}
    \hat{F}(k_y,i\omega_n)=
    \begin{pmatrix}
        F^{\uparrow\uparrow}(k_y,i\omega_n) & F^{\uparrow\downarrow}(k_y,i\omega_n) \\
        F^{\downarrow\uparrow}(k_y,i\omega_n) & F^{\downarrow\downarrow}(k_y,i\omega_n)
    \end{pmatrix},
\end{align}%
In this study, pair amplitude is classified as: even-frequency spin-singlet even-parity (ESE), even-frequency spin-triplet odd-parity (ETO), odd-frequency spin-singlet odd-parity (OSO)~\cite{BalatskyPRB1992,LinderRMP2019}, and odd-frequency spin-triplet even-parity (OTE)~\cite{berezinskii1974,KirkpatrickPRL1991,BergeretPRB2005,LinderRMP2019}.
Then we can calculate the pair amplitude at the edge ($j_x=j'_x=0$) as:
\begin{align}
    F^{\uparrow\downarrow}_\mathrm{ESE}(k_y,i\omega_{n})&=\frac{1}{4}[F^{\uparrow\downarrow}(k_y,i\omega_{n})-F^{\downarrow\uparrow}(k_y,i\omega_{n})\\
    &+F^{\uparrow\downarrow}(k_y,-i\omega_{n})-F^{\downarrow\uparrow}(k_y,-i\omega_{n})],\notag
\end{align}%
for ESE,
\begin{align}
    F^{\uparrow\downarrow}_\mathrm{ETO}(k_y,i\omega_{n})&=\frac{1}{4}[F^{\uparrow\downarrow}(k_y,i\omega_{n})+F^{\downarrow\uparrow}(k_y,i\omega_{n})\\
    &+F^{\uparrow\downarrow}(k_y,-i\omega_{n})+F^{\downarrow\uparrow}(k_y,-i\omega_{n})],\notag
\end{align}%
\begin{align}
    F^{\uparrow\uparrow}_\mathrm{ETO}(k_y,i\omega_{n})&=\frac{1}{2}[F^{\uparrow\uparrow}(k_y,i\omega_{n})+F^{\uparrow\uparrow}(k_y,-i\omega_{n})],
\end{align}%
\begin{align}
    F^{\downarrow\downarrow}_\mathrm{ETO}(k_y,i\omega_{n})&=\frac{1}{2}[F^{\downarrow\downarrow}(k_y,i\omega_{n})+F^{\downarrow\downarrow}(k_y,-i\omega_{n})],
\end{align}%
for ETO,
\begin{align}
    F^{\uparrow\downarrow}_\mathrm{OSO}(k_y,i\omega_{n})&=\frac{1}{4}[F^{\uparrow\downarrow}(k_y,i\omega_{n})-F^{\downarrow\uparrow}(k_y,i\omega_{n})\\
    &-F^{\uparrow\downarrow}(k_y,-i\omega_{n})+F^{\downarrow\uparrow}(k_y,-i\omega_{n})],\notag
\end{align}%
for OSO, and
\begin{align}
    F^{\uparrow\downarrow}_\mathrm{OTE}(k_y,i\omega_{n})&=\frac{1}{4}[F^{\uparrow\downarrow}(k_y,i\omega_{n})+F^{\downarrow\uparrow}(k_y,i\omega_{n})\\
    &-F^{\uparrow\downarrow}(k_y,-i\omega_{n})-F^{\downarrow\uparrow}(k_y,-i\omega_{n})],\notag
\end{align}%
\begin{align}
    F^{\uparrow\uparrow}_\mathrm{OTE}(k_y,i\omega_{n})&=\frac{1}{2}[F^{\uparrow\uparrow}(k_y,i\omega_{n})-F^{\uparrow\uparrow}(k_y,-i\omega_{n})],
\end{align}%
\begin{align}
    F^{\downarrow\downarrow}_\mathrm{OTE}(k_y,i\omega_{n})&=\frac{1}{2}[F^{\downarrow\downarrow}(k_y,i\omega_{n})-F^{\downarrow\downarrow}(k_y,-i\omega_{n})],
\end{align}%
for OTE pairings.

In the bulk, we can obtain the analytical solutions of the pair amplitude, see also in Appendix B.
When we select $t_{x}\ne0$ and $t_{y}=0$, ETO pairing is given by
\begin{align}
    F^{\uparrow\uparrow}_\mathrm{ETO}=F^{\downarrow\downarrow}_\mathrm{ETO}=\frac{2\Delta \eta Jt_x\sin{k_x}}{D_{x}},
\end{align}%
and OTE pairing is given by
\begin{align}
    F^{\uparrow\uparrow}_\mathrm{OTE}=F^{\downarrow\downarrow}_\mathrm{OTE}=\frac{2i\Delta \eta J \omega_{n}}{D_{x}},
\end{align}%
with $\eta=\pm 1$ and
\begin{align}
    D_{x}&=t^{4}_{x}\sin^{4}{k_x}+\Delta^{4}+\varepsilon^{4}(\bm{k})+J^{2}\notag\\
    &-2\varepsilon^{2}(\bm{k})J^{2}+2\varepsilon^{2}(\bm{k})\omega^{2}_{n}+2J^{2}\omega^{2}_{n}+\omega^{4}_{n}\notag\\
    &+2\Delta^{2}[\varepsilon^{2}(\bm{k})-J^{2}+\omega^{2}_{n}]\notag\\
    &+2t^{2}_{x}\sin^{2}{k_x}[\Delta^{2}-\varepsilon^{2}(\bm{k})+J^{2}+\omega^{2}_{n}].
\end{align}%
We obtain these equations as worked in Appendix B.
ETO pairing is induced by both nonzero $t_{x,y}$ and $J$; however, OTE pairing can be exhibited with nonzero $J$, and it is independent of $t_{x,y}$.

\begin{figure}[t!]
    \centering
    \includegraphics[width=8.5cm]{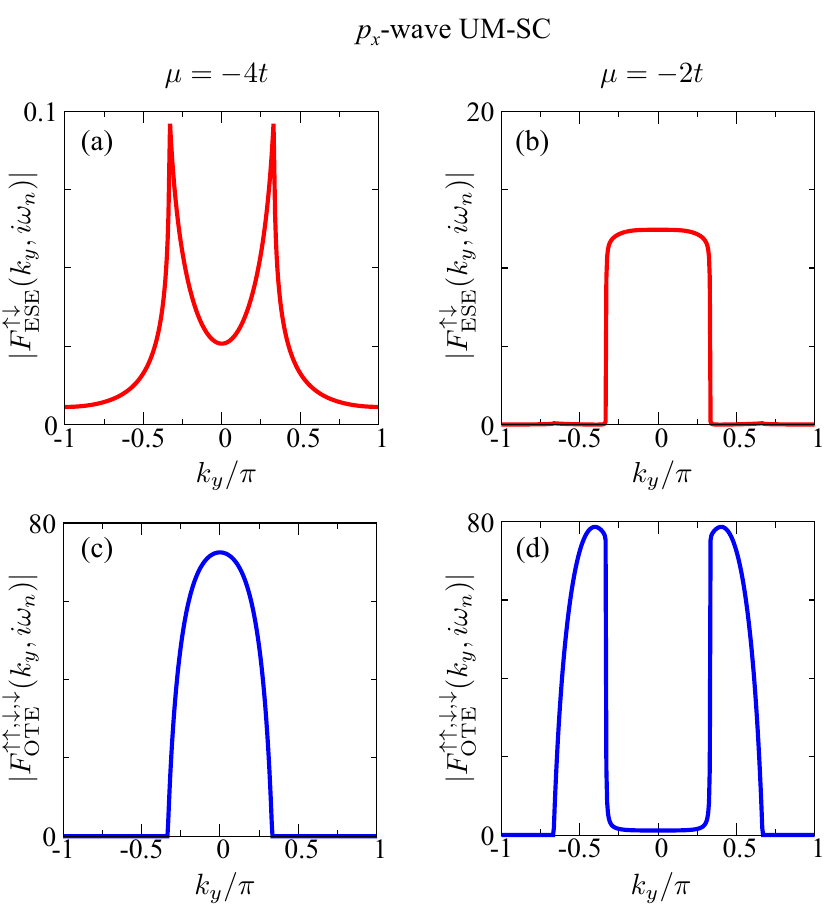}
    \caption{Absolute value of the pair amplitude for (a)(b) even-frequency spin-singlet even-parity (ESE) and (c)(d) odd-frequency spin-triplet even-parity (OTE) pairings at the [100] edge of PUM-SCs at $t_{x}=t$.
    We plot the pair amplitude at the lowest Matsubara frequency $\omega_{n}=\pi k_\mathrm{B}T$.
    The function of $|F^{\uparrow\uparrow}_\mathrm{OTE}|$ and $|F^{\downarrow\downarrow}_\mathrm{OTE}|$ for $k_y$ are the same.
    Parameters: $(t_{x},t_{y})=(t,0)$, $J=t$, $T_\mathrm{c}=0.01t$, and $T=0.025T_\mathrm{c}$.}
    \label{fig:4}
\end{figure}%
\begin{figure}[t!]
    \centering
    \includegraphics[width=8.5cm]{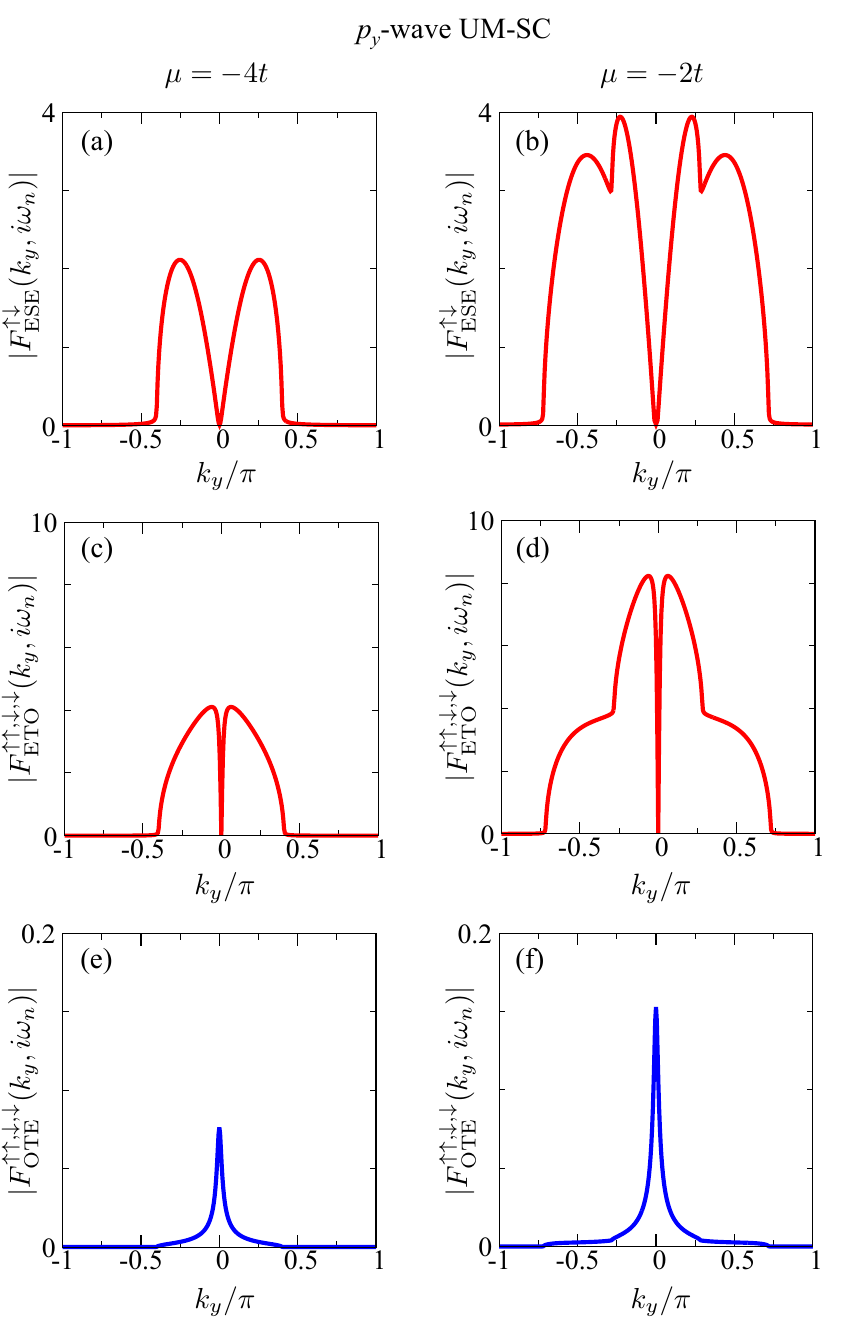}
    \caption{Absolute value of the pair amplitude for (a)(b) even-frequency spin-singlet even-parity (ESE), (c)(d) even-frequency spin-triplet odd-parity (ETO), and (e)(f) odd-frequency spin-triplet even-parity (OTE) pairings at the [100] edge of PUM-SCs at $t_{y}=t$.
    We plot the pair amplitude at the lowest Matsubara frequency $\omega_{n}=\pi k_\mathrm{B}T$.
    The function of $|F^{\uparrow\uparrow}_\mathrm{ETO}|$ and $|F^{\downarrow\downarrow}_\mathrm{ETO}|$ for $k_y$ are the same.
    Parameters: $(t_{x},t_{y})=(0,t)$, $J=t$, $T_\mathrm{c}=0.01t$, and $T=0.025T_\mathrm{c}$.}
    \label{fig:5}
\end{figure}%

While at the edge, we need to calculate the semi-infinite Green's function for $\omega_{n}=\pi k_\mathrm{B}T$ to evaluate the pair amplitude.
Calculating the pair amplitude for $\omega_{n}=\pi k_\mathrm{B}T$, we assume that the superconducting energy gap amplitude has the temperature $T$ dependence that obeys the Bardeen-Cooper-Schrieffer (BCS) theory: 
\begin{align}
    \Delta(T)=\Delta_{0}\tanh\left[1.74\sqrt{\frac{T_\mathrm{c}-T}{T}}\right], \label{DT_1}
\end{align}%
with the gap amplitude at zero temperature $\Delta_{0}=1.76T_\mathrm{c}$ and the critical temperature $T_\mathrm{c}=0.01t$.
In this section, we choose the temperature as $T=0.025T_\mathrm{c}$.

We plot the absolute value of the pair amplitude for ESE and OTE pairings at $(t_{x},t_{y})=(t,0)$ ($p_x$-wave UM) in Fig.~\ref{fig:4}.
Then we clarify the pair amplitude by the lowest Matsubara frequency $\omega_{n}=\pi k_\mathrm{B}T$.
In this case, ETO and OSO components vanish at the [100] edge owing to the nonlocal pairings along the $x$-direction.
As shown in Appendix B, ETO pairing with equal spins is dominant in the bulk.
At the [100] edge, at $\mu=-4t$ in Fig.~\ref{fig:4} (a) and Fig.~\ref{fig:4} (c), the OTE pairing $|F^{,\uparrow\uparrow,\downarrow\downarrow}_\mathrm{OTE}(k_y,i\omega_{n})|$ is around 800 times larger than ESE $|F^{\uparrow\downarrow}_\mathrm{ESE}(k_y,i\omega_{n})|$.
Then the OTE component is enhanced in the presence of zero-energy flat bands [Fig.~\ref{fig:2} (b)].
At $\mu=-2t$, ESE pairing becomes large [Fig.~\ref{fig:4} (b)] as compared with at $\mu=-4t$ [Fig.~\ref{fig:4} (a)], and nonzero values appear in the absence of zero-energy flat bands [Fig.~\ref{fig:2} (c)], while OTE pairing is also enhanced in the presence of zero-energy flat bands [Fig.~\ref{fig:2} (c)].

At $(t_{x},t_{y})=(0,t)$ ($p_y$-wave UM) in Fig.~\ref{fig:5}, we also obtain ESE, ETO, and OTE pairings at the [100] edge.
Because the ETO pairing is odd for $k_y$ in the bulk, as we worked in Appendix B, its component becomes nonzero at the edge [Fig.~\ref{fig:5} (c) and Fig.~\ref{fig:5} (d)].
As well as the $p_x$-wave UM-SC, ETO pairing is also dominant in the bulk, see also in Appendix B.
As compared with $(t_{x},t_{y})=(t,0)$ ($p_x$-wave UM-SC) in Fig.~\ref{fig:4}, ESE and ETO pairings are dominant.
Because we do not obtain zero-energy flat bands at $(t_{x},t_{y})=(0,t)$ ($p_y$-wave UM-SC), OTE pairing is smaller than other pair amplitudes~\ref{fig:5}.
These behaviors are consistent with the fact that the OTE state is enhanced at the edge in the presence of zero-energy flat bands.
However, ESE pairing does not vanish, even though $p$-wave superconductivity emerges in PUM-SCs.
Indeed, ETO pairing with equal spins is dominant in the bulk, and OTE pairing that is enhanced in the presence of zero-energy flat bands can also exist at the edge of PUM-SCs. 
Then, the magnitude of the ESE pairing at the [100] edge is suppressed in the presence of zero-energy flat bands.

\section{Josephson current in PUM-SCs in the high-transparency limit}

Next, we demonstrate the Josephson current in Josephson junctions with PUM-SCs.
In the present study, we assume the periodic boundary condition along the $y$-direction and semi-infinite SCs on both sides of the junctions.
Then we also consider two layers in the normal metal of the junction, see also Appendix C.
In the tight-binding model, the Josephson current is calculated by~\cite{KawaiPRB2017,FukayaPRB2020,fukaya2022npj,sakamori2023,fukaya2024}
\begin{align}
    I(\varphi)&=\frac{iek_\mathrm{B}T}{\hbar}\int dk_{y}\sum_{\omega_{n}}\mathrm{Tr'}[\tilde{t}^{\dagger}_\mathrm{N}\tilde{G}_{0,1}(k_y,i\omega_{n})\notag\\
    &-\tilde{t}_\mathrm{N}\tilde{G}_{1,0}(k_y,i\omega_{n})], \label{JC}
\end{align}%
where $T$ is the temperature, $\tilde{t}_\mathrm{N}$ is the nearest-neighbor hopping term in the normal metal, and $\tilde{G}_{0,1}(k_y,i\omega_{n})$ and $\tilde{G}_{1,0}(k_y,i\omega_{n})$ are the nonlocal Green's functions, see also in Appendix C.
We summarize the formulation of the Josephson current in Appendix C.

In this study, we have revealed that the current phase relation is described by the odd function of the Fourier series~\cite{GolubovJJReview}:
\begin{align}
    I(\varphi)=\sum_{m}I_{m}\sin(m\varphi),
\end{align}
with the phase difference $\varphi$.
We hereby focus on the high-transparency limit, with the higher harmonics components $I_{m}$~\cite{Kulik}. 
Then we choose the transparency amplitude as $t_\mathrm{int}=1.0$, see also in Appendix C.

\begin{figure}[t!]
    \centering
    \includegraphics[width=8.5cm]{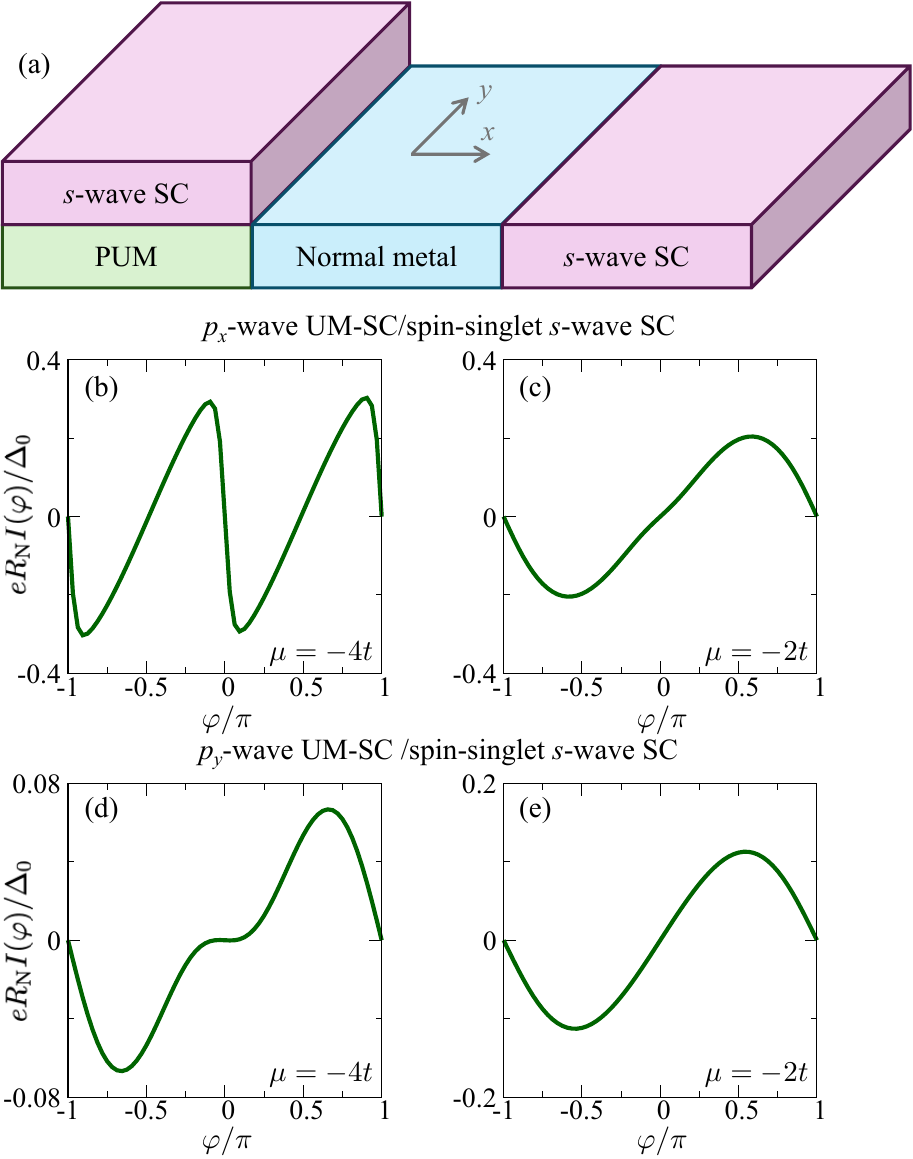}
    \caption{(a) Schematic image of PUM-SC/spin-singlet $s$-wave SC Josephson junctions.
    (b-e) Current phase relation in PUM-SC/spin-singlet $s$-wave SC Josephson junctions at (b)(c) $(t,t_{y})=(t,0)$ and (d)(e) $(t_{x},t_{y})=(0,t)$.
    We select the chemical potential as (b)(d) $\mu=-4t$ and (c)(e) $\mu=-2t$.
    Parameters: $\mu_\mathrm{N}=-3.5t$, $\mu_{s}=-1.5t$, $J=t$, $p=1$, $T_\mathrm{c}=0.01t$, $T=0.025T_\mathrm{c}$, and $t_\mathrm{int}=1$.}
    \label{fig:6}
\end{figure}%
\begin{figure}[t!]
    \centering
    \includegraphics[width=8.5cm]{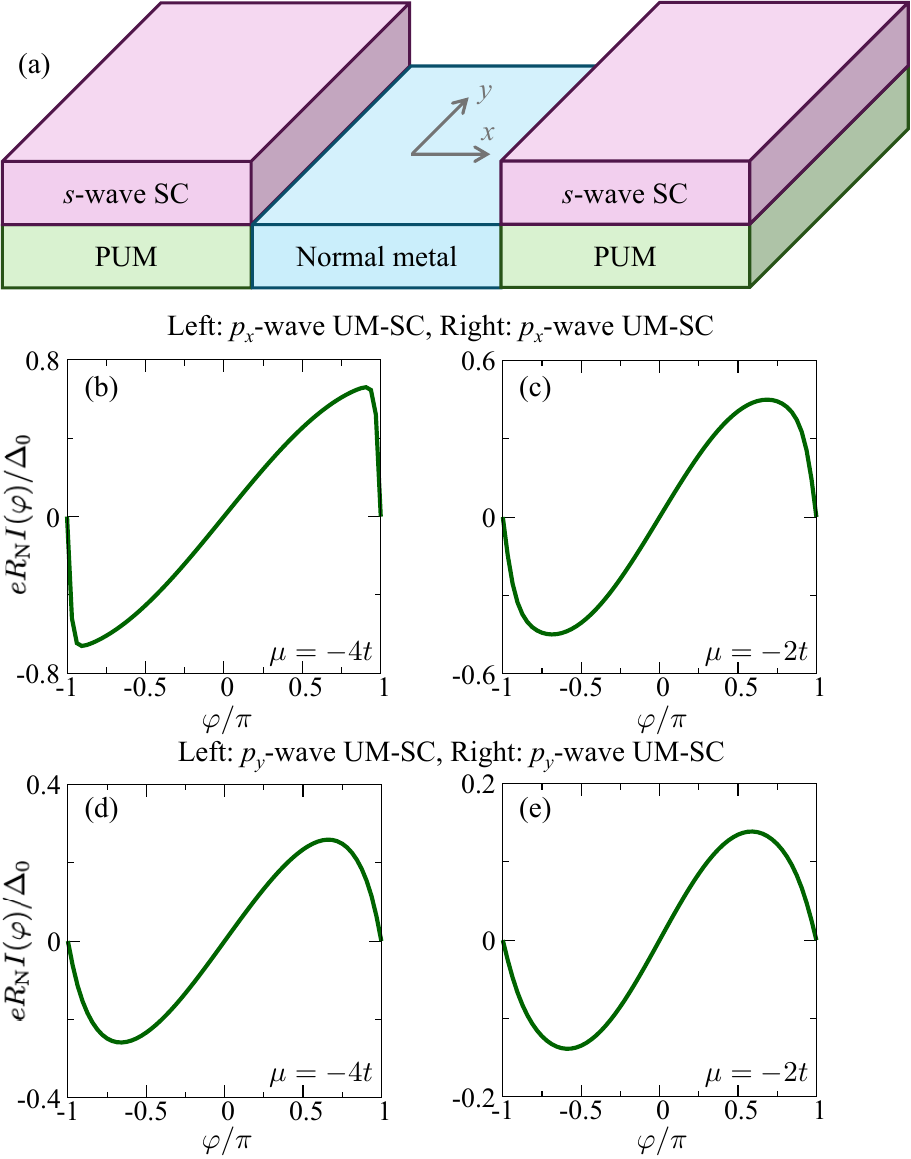}
    \caption{(a) Schematic image of PUM-SC/PUM-SC Josephson junctions.
    (b-e) Current phase relation in PUM-SC/PUM-SC Josephson junctions at (b)(c) $(t^\mathrm{L}_{x},t^\mathrm{L}_{y};t^\mathrm{R}_{x},t^\mathrm{R}_{y})=(t,0;t,0)$ and (d)(e) $(t^\mathrm{L}_{x},t^\mathrm{L}_{y};t^\mathrm{R}_{x},t^\mathrm{R}_{y})=(0,t;0,t)$.
    We choose the chemical potential as (b)(d) $\mu^\mathrm{L}=\mu^\mathrm{R}=-4t$ and (c)(e) $\mu^\mathrm{L}=\mu^\mathrm{R}=-2t$.
    Parameters: $\mu_\mathrm{N}=-3.5t$, $J^\mathrm{L}=J^\mathrm{R}=t$, $T_\mathrm{c}=0.01t$, $T=0.025T_\mathrm{c}$, and $t_\mathrm{int}=1$.}
    \label{fig:7}
\end{figure}%

\subsection{PUM-SC/spin-singlet $s$-wave SC Josephson junctions}

First of all, we study the Josephson current in PUM-SC/spin-singlet $s$-wave SC Josephson junctions shown in Fig.~\ref{fig:6} (a).
In the normal metal, the Hamiltonian is given by
\begin{align}
    \hat{H}_\mathrm{N}(\bm{k})&=[-\mu_\mathrm{N}-2t\cos{k_x}-2t\cos{k_y}]\hat{s}_{0},
\end{align}%
with the chemical potential in the normal metal $\mu_\mathrm{N}=-3.5t$.
In spin-singlet $s$-wave SCs, we assume the following Bogoliubov-de Gennes Hamiltonian:
\begin{align}
    \hat{H}^{s}_\mathrm{BdG}=
    \begin{pmatrix}
        \hat{H}_{s}(\bm{k}) & \hat{\Delta}_{s} \\
        \hat{\Delta}^{\dagger}_{s} & \hat{H}_{s}(\bm{k})
    \end{pmatrix},\label{BdG_eq_s_wave}
\end{align}%
\begin{align}
    \hat{H}_{s}(\bm{k})&=[-\mu_{s}-2t\cos{k_x}-2t\cos{k_y}]\hat{s}_{0},\\
    \hat{\Delta}_{s}&=\Delta_{s}(T) i\hat{s}_{2},
\end{align}%
\begin{align}
    \Delta_{s}(T)&=\bar\Delta_{s}\tanh\left[1.74\sqrt{\frac{T_\mathrm{c}-T}{T}}\right], \label{DT_s}
\end{align}%
with the chemical potential in spin-singlet $s$-wave SCs $\mu_{s}=-1.5t$ and the superconducting energy gap amplitude $\bar\Delta_{s}$.
Since the materials are different, we choose the distinct values of the chemical potentials in the spin-singlet $s$-wave SC $\mu_{s}$ and the normal metal $\mu_\mathrm{N}$.
Although the magnitude of the pair potential in PUM-SC $\bar\Delta_{p}$:
\begin{align}
    \Delta_{p}(T)&=\bar\Delta_{p}\tanh\left[1.74\sqrt{\frac{T_{\mathrm{c}}-T}{T}}\right], \label{DT_p}
\end{align}%
with the gap amplitude at zero temperature $\bar\Delta_{p}=p\Delta_{0}$, is smaller than that in spin-singlet $s$-wave SCs $\bar\Delta_{s}$~\cite{McMillanPR1968}, here, we choose $\bar\Delta_{p}=\Delta_{0}$ ($p=1$) and $\bar\Delta_{s}=\Delta_{0}$ for simplicity.
The current phase relations do not change qualitatively even though we set the smaller $\bar\Delta_{s}$ than $\Delta_{0}$ at high transparency, as shown in Appendix D.

For the $p_x$-wave UM-SC at $(t_{x},t_{y})=(t,0)$, we plot the current phase relation at Fig.~\ref{fig:6} (b) $\mu=-4t$ and Fig.~\ref{fig:6} (c) $\mu=-2t$.
Although the $I_{2}$ term is dominant at $\mu=-4t$ [Fig.~\ref{fig:6} (b)], the $I_{1}$ component remains.
It originates from the ESE pair amplitude at the [100] edge [Fig.~\ref{fig:4} (a)], even though we obtain zero-energy flat bands caused by $p_x$-wave superconductivity at $\mu=-4t$ as shown in Fig.~\ref{fig:2} (b).
Because this ESE pair amplitude on the left side is coupled to the spin-singlet $s$-wave superconductivity on the right side by the first order, the $I_{1}$ term does not disappear in Fig.~\ref{fig:6} (b).
In $p_x$-wave SC/spin-singlet $s$-wave SC Josephson junctions, the current phase relation is described by $I_{2}\sin(2\varphi)$, and $I_{1}$ vanishes~\cite{AsanoPRB2003}.
It indicates that PUM-SC/spin-singlet $s$-wave SC Josephson junctions are not equivalent to $p_x$-wave SC/$s$-wave SC ones.
As well as at $\mu=-2t$ [Fig.~\ref{fig:6} (c)], $I_{1}$ becomes nonzero and it is dominant as compared with higher $I_{m\ge2}$.
Since the amplitude of the ESE component at $\mu=-2t$ [Fig.~\ref{fig:4} (b)] is 200 times larger than that at $\mu=-4t$ [Fig.~\ref{fig:4} (a)], the configuration of $I_{1}$ is also enhanced.
We note that $\pi$-junctions~\cite{Pugach10} do not appear due to the conventional spin-singlet $s$-wave coupling on both sides of the junctions.
Likewise, for the $p_y$-wave UM-SC case at $(t_{x},t_{y})=(0,t)$, the $I_{1}$ component is also dominant at $\mu=-4t$ [Fig.~\ref{fig:6} (d)] and $\mu=-2t$ [Fig.~\ref{fig:6} (e)] owing to the larger ESE pair amplitude [Fig.~\ref{fig:5} (a) and Fig.~\ref{fig:5} (b)].
It is also noted that the current phase relations in Fig.~\ref{fig:6} (b) and Fig.~\ref{fig:6} (d) are $\varphi$-junctions, where the free energy minima of the junctions are located $\pm\varphi$ with degeneracy.
Hence, the conventional current phase relation with $I_{1}>0$ mainly emerges in PUM-SC/spin-singlet $s$-wave Josephson junctions.
It indicates that the current phase relation of PUM-SC/spin-singlet $s$-wave SC Josephson junctions does not coincide with that of $p$-wave SC/$s$-wave SC Josephson junctions, even though $p$-wave superconductivity is realized in PUM-SCs.
In Appendix D, we also show the current phase relations when changing the magnitude of the pair potential in PUM-SC at $p=0.1$, and the qualitative results of the current phase relations do not change.

\begin{figure}[t!]
    \centering
    \includegraphics[width=8.5cm]{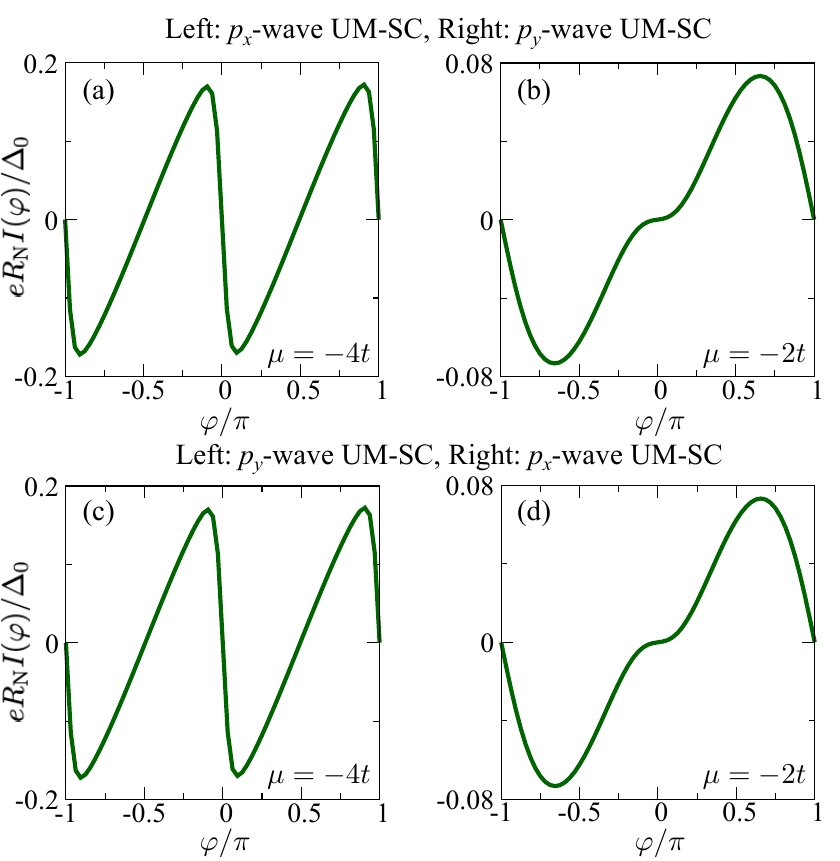}
    \caption{Current phase relation in PUM-SC/PUM-SC Josephson junctions at (a)(b) $(t^\mathrm{L}_{x},t^\mathrm{L}_{y};t^\mathrm{R}_{x},t^\mathrm{R}_{y})=(t,0;0,t)$ and (c)(d) $(t^\mathrm{L}_{x},t^\mathrm{L}_{y};t^\mathrm{R}_{x},t^\mathrm{R}_{y})=(0,t;t,0)$.
    We choose the chemical potential as (a)(c) $\mu^\mathrm{L}=\mu^\mathrm{R}=-4t$ and (b)(d) $\mu^\mathrm{L}=\mu^\mathrm{R}=-2t$.
    Parameters: $\mu_\mathrm{N}=-3.5t$, $J=t$, $T_\mathrm{c}=0.01t$, $T=0.025T_\mathrm{c}$, and $t_\mathrm{int}=1$.}
    \label{fig:8}
\end{figure}%

\subsection{PUM-SC/PUM-SC Josephson junctions}

Next, we study the current phase relation in PUM-SC/PUM-SC Josephson junctions [Fig.~\ref{fig:7} (a)].
Then we introduce the parameters on the left side: $(\mu^\mathrm{L},t^\mathrm{L}_{x},t^\mathrm{L}_{y},J^\mathrm{L})$ and the right side: $(\mu^\mathrm{R},t^\mathrm{R}_{x},t^\mathrm{R}_{y},J^\mathrm{R})$, respectively.
For simplicity, we choose $\mu=\mu^\mathrm{L}=\mu^\mathrm{R}$, $J^\mathrm{L}=J^\mathrm{R}=t$, and the same critical temperature $T_\mathrm{c}$ and the amplitude of the pair potential $\Delta(T)$ in Eq.\ (\ref{DT_1}) in Josephson junctions.

At $(t^\mathrm{L}_{x},t^\mathrm{L}_{y};t^\mathrm{R}_{x},t^\mathrm{R}_{y})=(t,0;t,0)$ (Left: $p_x$-wave UM-SC, Right: $p_x$-wave UM-SC), for $\mu=-4t$, the current phase relation becomes maximum around $\varphi=\pm\pi$ [Fig.~\ref{fig:7} (b)].
Then, the skewness of the current phase relation becomes remarkable in the presence of zero-energy flat bands.
This behavior is similar to that known in $p_x$-wave SC/$p_x$-wave SC Josephson junctions~\cite{TanakaPRB1999,kwon2004fractional,AsanoPRL2006}.
However, for $\mu=-2t$ shown in Fig.~\ref{fig:7} (c), the skewness effect in the current phase relation is suppressed.
Although the effective transparency for $\mu=-2t$ is distinct from that for $\mu=-4t$, this behavior is caused by the smaller regime of zero-energy flat bands formed in the momentum space [Fig.~\ref{fig:2} (c)] compared with Fig.~\ref{fig:2} (b).
In Fig.~\ref{fig:7} (d) for $\mu=-4t$ and Fig.~\ref{fig:7} (e) for $\mu=-2t$, at $(t^\mathrm{L}_{x},t^\mathrm{L}_{y};t^\mathrm{R}_{x},t^\mathrm{R}_{y})=(0,t;0,t)$ (Left: $p_y$-wave UM-SC, Right: $p_y$-wave UM-SC), we obtain the $I_1$-dominant current phase relations.
We also calculate the current phase relation at $(t^\mathrm{L}_{x},t^\mathrm{L}_{y};t^\mathrm{R}_{x},t^\mathrm{R}_{y})=(t,0;0,t)$ (Left: $p_x$-wave UM-SC, Right: $p_y$-wave UM-SC) and $(t^\mathrm{L}_{x},t^\mathrm{L}_{y};t^\mathrm{R}_{x},t^\mathrm{R}_{y})=(0,t;t,0)$ (Left: $p_y$-wave UM-SC, Right: $p_x$-wave UM-SC) shown in Fig.~\ref{fig:8}.
Because induced spin-triplet pairings with different mirror symmetries in the $zx$-plane are coupled within the second order, the current phase relation does not change by the exchange of the left and right sides [Fig.~\ref{fig:8}].
For $\mu=-4t$, $I_2$-dominant behavior is obtained, and then $I_1$ is nonzero [Fig.~\ref{fig:8} (a) and Fig.~\ref{fig:8} (c)].
In Fig.~\ref{fig:8} (b) and Fig.~\ref{fig:8} (d), for $\mu=-2t$, $I_{1}$ is comparable with $I_{2}$.
These behaviors correspond to the $\varphi$-junctions~\cite{TKdJJ96,tanaka971}.
Indeed, the coupling of the spin-triplet $p$-wave pair amplitude on both sides of the junctions contributes to the current phase relation; however, spin-singlet pairings also play a role in the current phase relation of PUM-SC/PUM-SC Josephson junctions.

\begin{figure}[t!]
    \centering
    \includegraphics[width=8.5cm]{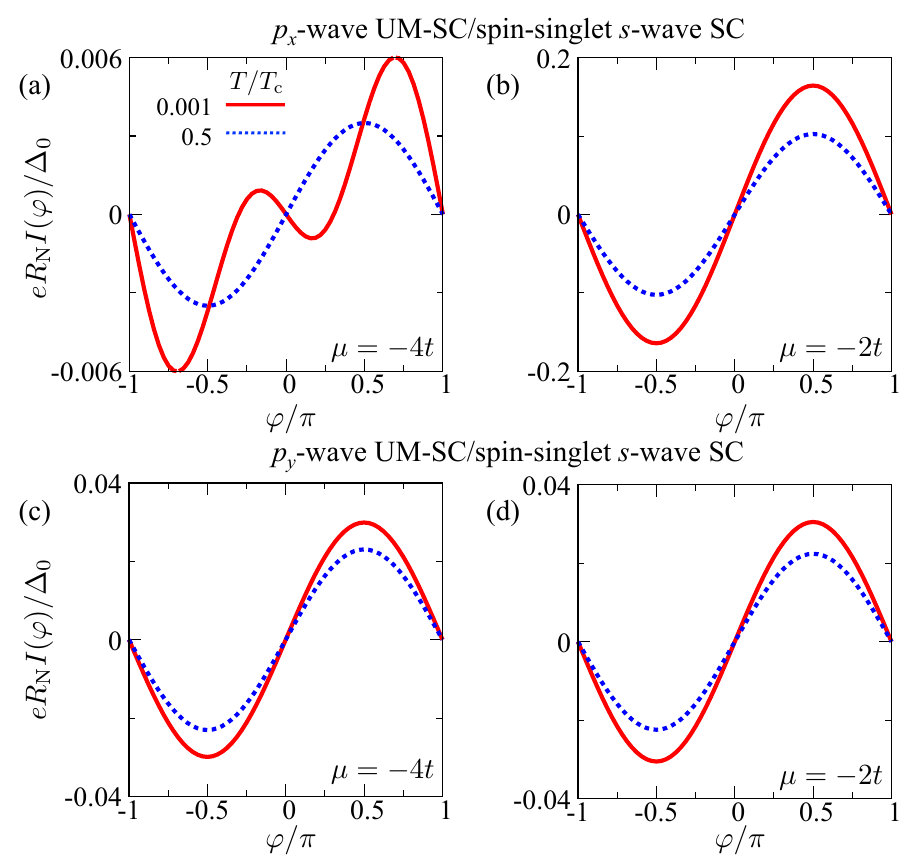}
    \caption{Current phase relation in PUM-SC/spin-singlet $s$-wave SC Josephson junctions at (a)(b) $(t_{x},t_{y})=(t,0)$ and (c)(d) $(t_{x},t_{y})=(0,t)$.
    We select the chemical potential as (a)(c) $\mu=-4t$ and (b)(d) $\mu=-2t$.
    Red-solid and blue-dotted lines correspond to the temperature $T=0.001T_\mathrm{c}$ and $T=0.5T_\mathrm{c}$, respectively.
    Parameters: $\mu_\mathrm{N}=-3.5t$, $\mu_{s}=-1.5t$, $J=t$, $p=1$, $T_\mathrm{c}=0.01t$, and $t_\mathrm{int}=0.1$.}
    \label{fig:9}
\end{figure}%
\begin{figure}[t!]
    \centering
    \includegraphics[width=8.5cm]{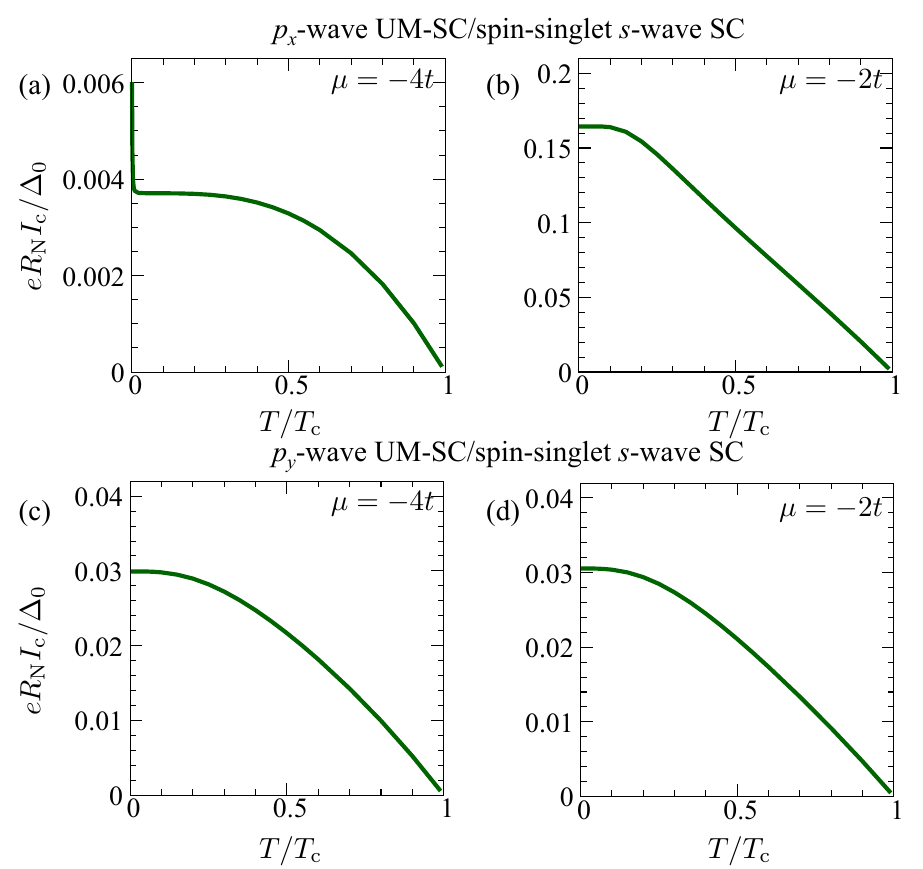}
    \caption{Temperature dependence of the maximum Josephson current $I_\mathrm{c}$ in PUM-SC/spin-singlet $s$-wave SC Josephson junctions at (a) $(t_{x},t_{y},\mu)=(t,0,-4t)$, (b) $(t_{x},t_{y},\mu)=(t,0,-2t)$, (c) $(t_{x},t_{y},\mu)=(0,t,-4t)$, and (d) $(t_{x},t_{y},\mu)=(0,t,-2t)$.
    Parameters: $\mu_\mathrm{N}=-3.5t$, $\mu_{s}=-1.5t$, $J=t$, $T_\mathrm{c}=0.01t$, $p=1$, $\Delta_{0}=1.76T_\mathrm{c}$, and $t_\mathrm{int}=0.1$.}
    \label{fig:10}
\end{figure}%

\section{Temperature dependence of the maximum Josephson current in PUM-SCs in the low-transparency limit}

In this section, we calculate not only the current phase relation but also the temperature dependence of the maximum Josephson current in PUM-SCs.
Then we focus on the low transparency, and we select the transparency amplitude as $t_\mathrm{int}=0.1$ (see also in Appendix C).

In PUM-SC/spin-singlet $s$-wave Josephson junctions, we plot the current phase relations at $T=0.001T_\mathrm{c}$ (red-solid) and $T=0.5T_\mathrm{c}$ (blue-dotted lines) as shown in Fig.~\ref{fig:9}.
Then, we set the temperature dependence of the magnitude of the pair potential as Eq.\ (\ref{DT_p}) at $p=1$ in PUM-SC and Eq.\ (\ref{DT_s}) in spin-singlet $s$-wave SCs. 
At $T=0.001T_\mathrm{c}$, for $(t_{x},t_{y},\mu)=(t,0,-4t)$ ($p_x$-wave UM-SC) [Fig.~\ref{fig:9} (a)], the $\varphi$-junction appears owing to the strong resonance of zero-energy flat bands in $p_x$-wave UM-SCs~\cite{BarashPRL199,tanaka971}.
In Appendix D, for $(t_{x},t_{y},\mu)=(t,0,-4t)$, we confirm the current phase relations at $p=0.1$.
The non-sinusoidal feature disappears at $T=0.001T_\mathrm{c}$.
Due to the enhancement of higher harmonics in the current phase relation $I_{m\ge2}$ at low temperature, the maximum Josephson current $I_\mathrm{c}$ is also enhanced, although $I_\mathrm{c}$ is once saturated [Fig.~\ref{fig:10} (a)].
Because the region of zero-energy flat bands in the momentum space for $\mu=-2t$ is smaller than that for $\mu=-4t$ [Fig.~\ref{fig:2} (b)(c)], the $I_{1}\sin\varphi$ dependence is mainly dominant for $(t_{x},t_{y},\mu)=(t,0,-2t)$ ($p_x$-wave UM-SC) [Fig.~\ref{fig:9} (b)]. 
While, for $(t_{x},t_{y},\mu)=(0,t,-4t)$ [Fig.~\ref{fig:9} (c)] and $(t_{x},t_{y},\mu)=(0,t,-2t)$ ($p_y$-wave UM-SC) [Fig.~\ref{fig:9} (d)], we obtain $I_{1}\sin\varphi$ behaviors, and it is known that the current phase relation in conventional $s$-wave SC junctions is described by $I(\varphi)=I_{1}\sin{\varphi}$ (Ambegaokar-Baratoff behavior) in the low-transparency limit~\cite{Ambegaokar}, as well as at $T=0.50T_\mathrm{c}$.
The current phase relations do not change qualitatively at $(t_{x},t_{y},\mu)=(t,0,-2t)$, $(0,t,-4t)$, and $(t_{x},t_{y},\mu)=(0,t,-2t)$, see Appendix D.
The maximum Josephson current $I_\mathrm{c}$ for $(t_{x},t_{y},\mu)=(t,0,-2t)$ ($p_x$-wave UM-SC) [Fig.~\ref{fig:10} (b)], and $(t_{x},t_{y},\mu)=(0,t,-4t)$ [Fig.~\ref{fig:10} (c)] and $(t_{x},t_{y},\mu)=(0,t,-2t)$ ($p_y$-wave UM-SC) [Fig.~\ref{fig:10} (d)] is saturated at low temperature, and this behavior is independent of both the chemical potential $\mu=-4t,-2t$ and the spin-dependent hopping terms $t_{x,y}$.
As we showed in Fig.~\ref{fig:9}, the $I_{1}$ component does not vanish owing to the coupling of even-frequency spin-singlet even-parity pairings on both sides of the junctions.
In the case when the current phase relation becomes sinusoidal behavior that is obtained by Ambegaokar and Baratoff~\cite{Ambegaokar}, the maximum Josephson current is saturated at low temperature. 
As a result, in PUM-SC/spin-singlet $s$-wave junctions, we do not obtain the same temperature dependence as in 
spin-triplet $p$-wave SC/spin-singlet $s$-wave SC junctions without PUM.
Since the current phase relations depend on $\bar\Delta_{p}$, the temperature dependence is influenced by the choice of $\bar{\Delta}_{p}$.

\begin{figure}[t!]
    \centering
    \includegraphics[width=8.5cm]{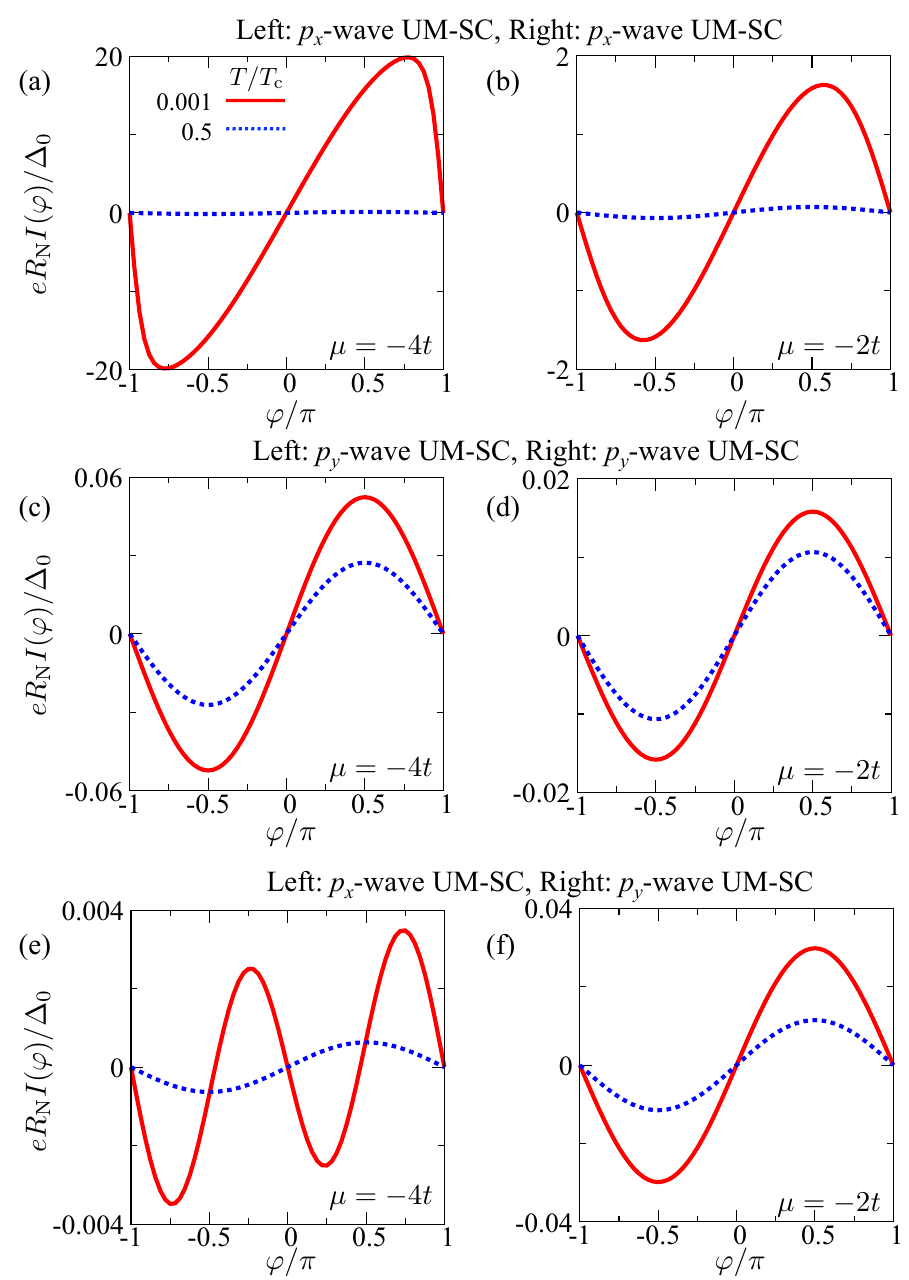}
    \caption{Current phase relation in PUM-SC/PUM-SC Josephson junctions at (a)(b) $(t^\mathrm{L}_{x},t^\mathrm{L}_{y};t^\mathrm{R}_{x},t^\mathrm{R}_{y})=(t,0;t,0)$, (c)(d) $(t^\mathrm{L}_{x},t^\mathrm{L}_{y};t^\mathrm{R}_{x},t^\mathrm{R}_{y})=(0,t;0,t)$, and (c)(d) $(t^\mathrm{L}_{x},t^\mathrm{L}_{y};t^\mathrm{R}_{x},t^\mathrm{R}_{y})=(t,0;0,t)$.
    We choose the chemical potential as (a)(c)(e) $\mu=\mu^\mathrm{L}=\mu^\mathrm{R}=-4t$ and (b)(d)(f) $\mu=\mu^\mathrm{L}=\mu^\mathrm{R}=-2t$.
    Red-solid and blue-dotted lines correspond to the temperature $T=0.001T_\mathrm{c}$ and $T=0.5T_\mathrm{c}$, respectively.
    Parameters: $\mu_\mathrm{N}=-3.5t$, $J^\mathrm{L}=J^\mathrm{R}=t$, $T_\mathrm{c}=0.01t$, and $t_\mathrm{int}=0.1$.}
    \label{fig:11}
\end{figure}%
\begin{figure}[t!]
    \centering
    \includegraphics[width=8.5cm]{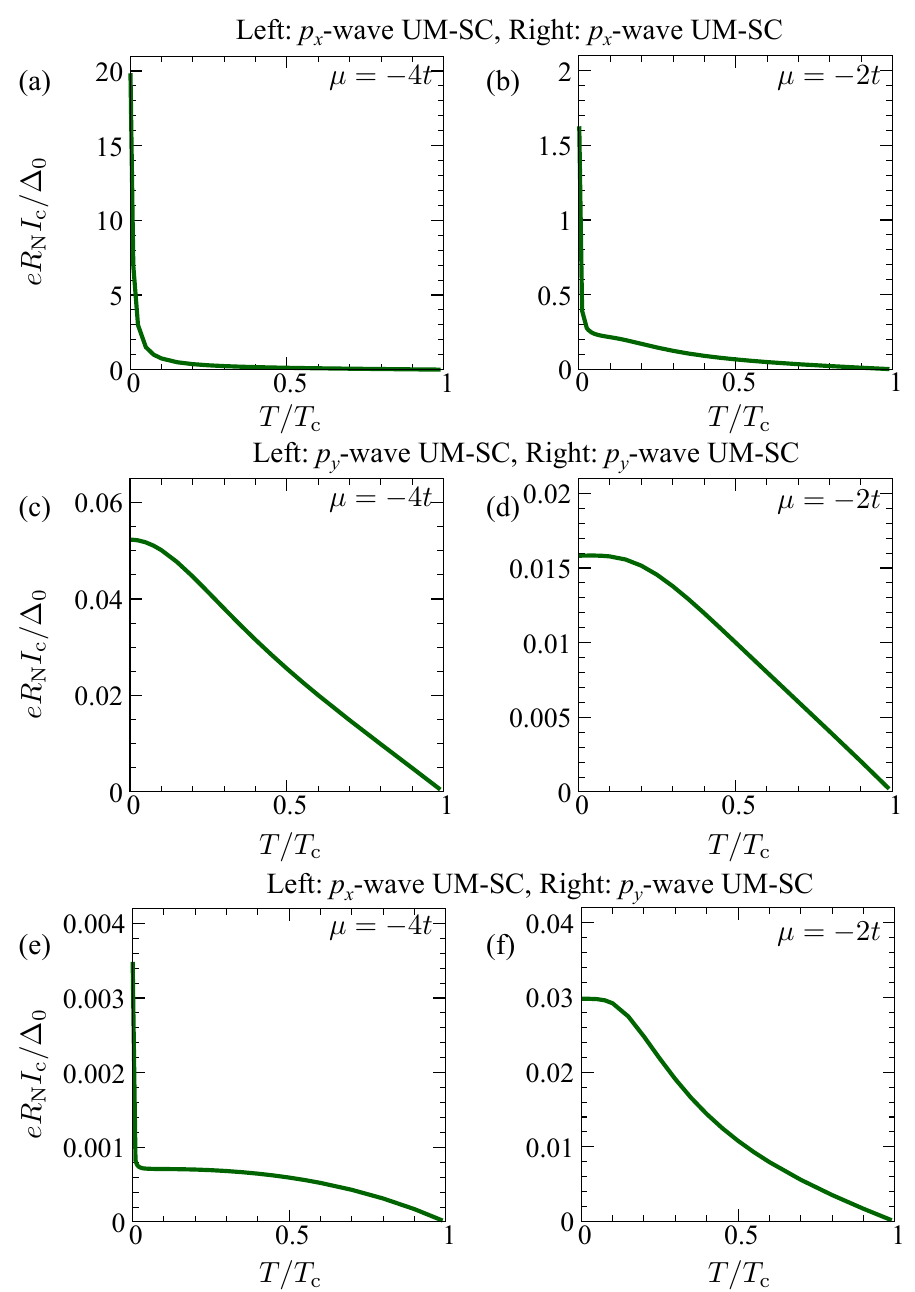}
    \caption{Temperature dependence of the maximum Josephson current $I_\mathrm{c}$ in PUM-SC/PUM-SC Josephson junctions at (a)(b) $(t^\mathrm{L}_{x},t^\mathrm{L}_{y};t^\mathrm{R}_{x},t^\mathrm{R}_{y})=(t,0;t,0)$, (c)(d) $(t^\mathrm{L}_{x},t^\mathrm{L}_{y};t^\mathrm{R}_{x},t^\mathrm{R}_{y})=(0,t;0,t)$, and (e)(f) $(t^\mathrm{L}_{x},t^\mathrm{L}_{y};t^\mathrm{R}_{x},t^\mathrm{R}_{y})=(t,0;0,t)$.
    We choose the chemical potential as (a)(c)(e) $\mu=\mu^\mathrm{L}=\mu^\mathrm{R}=-4t$ and (b)(d)(f) $\mu=\mu^\mathrm{L}=\mu^\mathrm{R}=-2t$.
    Parameters: $\mu_\mathrm{N}=-3.5t$, $J=t$, $T_\mathrm{c}=0.01t$, $\Delta_{0}=1.76T_\mathrm{c}$, and $t_\mathrm{int}=0.1$.}
    \label{fig:12}
\end{figure}%

Next, we discuss the current phase relation and temperature dependence of Josephson current in PUM-SC/PUM-SC junctions.
In the current phase relation shown in Fig.~\ref{fig:11}, at $T=0.5T_\mathrm{c}$ (blue-dotted line), we obtain $I_{1}\sin{\varphi}$ behavior in all cases.
Then, we select the temperature dependence of the magnitude of the pair potential as Eq.\ (\ref{DT_1}). 
At $T=0.001T_\mathrm{c}$ (red-solid line), for $(t^\mathrm{L}_{x},t^\mathrm{L}_{y};t^\mathrm{R}_{x},t^\mathrm{R}_{y})=(t,0;t,0)$ (Left: $p_x$-wave UM-SC, Right: $p_x$-wave UM-SC) with the chemical potential $\mu=-4t,-2t$, the current phase relation has a  skewness [Fig.~\ref{fig:11} (a) and Fig.~\ref{fig:11}(b)].
We note that the skewness at $\mu=-2t$ is not remarkable due to the smaller region of zero-energy flat bands in the momentum space as compared with that for $\mu=-4t$ [Fig.~\ref{fig:2} (b) and Fig.~\ref{fig:2} (c)].
The maximum Josephson current $I_\mathrm{c}$, in Fig.~\ref{fig:12} (a) and Fig.~\ref{fig:12} (b), at $(t^\mathrm{L}_{x},t^\mathrm{L}_{y};t^\mathrm{R}_{x},t^\mathrm{R}_{y})=(t,0;t,0)$ (Left: $p_x$-wave UM-SC, Right: $p_x$-wave UM-SC), is enhanced at low temperature for $\mu=-4t,-2t$.
This behavior originates from the resonance of zero-energy flat bands [Fig.~\ref{fig:2} (b) and Fig.~\ref{fig:2} (c)] on both sides of the junctions, similar to the case in $d$-wave SC junctions~\cite{BarashPRL199,tanaka971}.
For $(t^\mathrm{L}_{x},t^\mathrm{L}_{y};t^\mathrm{R}_{x},t^\mathrm{R}_{y})=(t,0;0,t)$ (Left: $p_x$-wave UM-SC, Right: $p_y$-wave UM-SC) with the chemical potential $\mu=-4t$, the $\varphi$-junction emerges [Fig.~\ref{fig:11} (e)], as we showed in $p_x$-wave UM-SC/$p_y$-wave UM-SC junctions for $\mu=-4t$ at the high-transparency limit [Fig.~\ref{fig:8} (a)].
Then, once $I_\mathrm{c}$ is saturated, it is enhanced, with the decrease of the temperature $T$ [Fig.~\ref{fig:12} (e)].
This enhancement is caused by the resonance of zero-energy flat bands in $p_x$-wave UM-SCs of the left side~\cite{BarashPRL199,tanaka971}.
In the other cases, at $(t^\mathrm{L}_{x},t^\mathrm{L}_{y};t^\mathrm{R}_{x},t^\mathrm{R}_{y})=(0,t;0,t)$ (Left: $p_y$-wave UM-SC, Right: $p_y$-wave UM-SC) for $\mu=-4t,-2t$ and at $(t^\mathrm{L}_{x},t^\mathrm{L}_{y};t^\mathrm{R}_{x},t^\mathrm{R}_{y})=(t,0;0,t)$ (Left: $p_x$-wave UM-SC, Right: $p_y$-wave UM-SC) for $\mu=-2t$, we obtain $I_{1}\sin\varphi$ dependence in [Figs.~\ref{fig:11}(c)(d)(f)].
At $(t^\mathrm{L}_{x},t^\mathrm{L}_{y};t^\mathrm{R}_{x},t^\mathrm{R}_{y})=(0,t;0,t)$ (Left: $p_y$-wave UM-SC, Right: $p_y$-wave UM-SC) for $\mu=-4t,-2t$, $I_\mathrm{c}$ is saturated at low temperature [Fig.~\ref{fig:12} (c) and Fig.~\ref{fig:12} (d)], in the absence of zero-energy flat bands [Fig.~\ref{fig:3} (b) and Fig.~\ref{fig:3} (c)].
Likewise, at $(t^\mathrm{L}_{x},t^\mathrm{L}_{y};t^\mathrm{R}_{x},t^\mathrm{R}_{y})=(t,0;0,t)$ (Left: $p_x$-wave UM-SC, Right: $p_y$-wave UM-SC) for $\mu=-2t$, the behavior of $I_\mathrm{c}$ changes [Fig.~\ref{fig:12} (f)], as compared with Fig.~\ref{fig:12} (e).
Since the region of zero-energy flat bands in the momentum space at $\mu=-2t$ is narrower than that at $\mu=-4t$, the enhancement of $I_\mathrm{c}$ does not occur in Fig.~\ref{fig:12} (f).
Hence, the spin-triplet $p$-wave pair amplitude in PUM-SCs can contribute to the current phase relation of the Josephson current with temperature, as well as the existence of zero-energy flat bands.

\section{Summary and Conclusion}

We have studied SDOS at the [100] edge in PUM-SCs.
We have shown that zero-energy flat bands can exist owing to the noncollinear spin structure in the $zx$-plane, and then $p$-wave magnetic order can generate $p$-wave superconductivity.
We note that noncollinear spin structure in the $xy$-plane can also lead to zero-energy flat bands at the [100] edge.
Similar behavior has already been theoretically proposed in a one-dimensional spin helix/$s$-wave SC hybrid system~\cite{KlinovajaPRL2013} and magnetic chain~\cite{EbisuPRB2015}.
Indeed, our results are qualitatively consistent with these previous works~\cite{BrauneckerPRB2010,NakosaiPRB2013,MartinPRB2012,NadjPRB2013,EbisuPRB2015}.

We have also investigated the Josephson current in superconducting junctions with PUM-SCs.
In the high-transparency limit, in PUM-SC/spin-singlet $s$-wave SC Josephson junctions, even though $p$-wave superconductivity emerges in PUM-SCs, the first harmonics term $I_{1}$ does not vanish owing to the coupling of spin-singlet $s$-wave superconductivity on both sides of the junctions.
Then, since the amplitude of $I_{1}$ can be suppressed, $\varphi$-junctions can be realized.
In PUM-SC/PUM-SC Josephson junctions, because even-frequency spin-triplet odd-parity pair amplitudes can be coupled on both sides of the junctions, the current phase relations of the Josephson current can be determined by the edge states of PUM-SCs.
In the low transparency limit, the maximum Josephson current can be saturated at low temperature in PUM-SC/spin-singlet $s$-wave SC junctions.
It originates from the coupling of even-frequency spin-singlet even-parity pairings on both sides of the junctions.
In PUM-SC/PUM-SC Josephson junctions, the resulting Josephson current $I_\mathrm{c}$ clearly depends on the absence or the presence of the edge states.
Our findings indicate that the nature of the $p$-wave superconductivity in PUM-SC is not visible in some Josephson current, owing to the remaining spin-singlet $s$-wave pairing. 
However, zero-energy flat bands stemming from the spin-triplet $p$-wave state can contribute to the Josephson current, and this behavior can be tunable by the chemical potential, which determines the generation of zero-energy flat bands.
Our findings indicate that PUM-SCs can be regarded as $s+p$-wave superconductors.

In this work, we have demonstrated the specific situation: $p_x$-wave and $p_y$-wave UM, and the chemical potential at $\mu=-4t,-2t$.
The presence of zero-energy flat bands caused by the spin-triplet $p_x$-wave superconductivity mainly plays a role in the current phase relation with higher harmonics and skewness.
It indicates that the profile of the current phase relation is determined by the presence or absence of zero-energy flat bands stemming from spin-triplet $p_x$-wave superconductivity.
Indeed, our findings can be applied to generic situations in distinct angles of the PUM Fermi surface tuned by $(t_{x},t_{y})$ and the chemical potential in PUMs.

In Ref.\ \cite{NagaePRB2025,SunZTPRB2025}, zero-energy flat bands have also been reported by using the other PUM model~\cite{hellenes2024P}.
Our results have been demonstrated in the PUM model suggested in Ref.\ \cite{BrekkePRL2024}, and the qualitative results are the same as Ref.\ \cite{NagaePRB2025,SunZTPRB2025}.
Although the model Hamiltonian and the electronic structure in Refs.~\cite{hellenes2024P,LeePRL2025} are different, the noncollinear spin structures that describe the $p$-wave unconventional magnets are essentially the same.
It means that nodal structures shown in Figs.~\ref{fig:2} (b)(c) and Figs.~\ref{fig:3} (b)(c) are defined by winding numbers, and these zero-energy flat bands are also topologically protected~\cite{SatoPRB2011}, and we have obtained a similar behavior, even though we have used a simpler model~\cite{BrekkePRL2024}.
Hence, the model in Ref.~\cite{BrekkePRL2024} can also capture the essential physics caused by noncollinear spin structures of $p$-wave unconventional magnetism, and our conclusion does not change qualitatively, even though we choose a different model Hamiltonian.
In addition, we expect that both the even-frequency spin-singlet even-parity pair amplitude also coexist with spin-triplet $p$-wave superconductivity in Ref.\ \cite{NagaePRB2025,SunZTPRB2025}.
Indeed, the Josephson current in our results can also be demonstrated by using the other PUM model~\cite{hellenes2024P}.

As a future perspective, PUM-SC/PUM/spin-singlet $s$-wave SC and PUM-SC/PUM/PUM-SC Josephson junctions can be considered to access realistic situations in experiments.
Then, since even-frequency spin-singlet even-parity pairings affect the Josephson current, as we demonstrated, the research of $s$-wave SC/PUM/$s$-wave SC junctions is needed to unveil the role of $p$-wave magnetic order.
In Ref.\ \cite{fukaya2024}, the Josephson currents were also obtained in $s$-wave SC/PUM/$s$-wave junctions with the simplified PUM model.
Thus, adopting effective PUM models~\cite{BrekkePRL2024,hellenes2024P} can support an experimental perspective in Josephson junctions with PUM.


\section{Acknowledgments}

Y.\ F.\ acknowledges financial support from the Sumitomo Foundation and the calculation support from Okayama University and JSPS with Grants-in-Aid for Scientific Research (KAKENHI Grants No.\ 26K17096).   
K.\ Y., Y.\ F., and Y.\ T.\ acknowledge financial support from JSPS with Grants-in-Aid for Scientific Research (KAKENHI Grants No.\ 25K07203).  
Y. T. acknowledges financial support from JSPS with Grants-in-Aid for Scientific Research (KAKENHI Grants Nos.\ 23K17668, 24K00583, 24K00556,
24K00578, 25H00609, and 25H00613). 
We thank B.\ Lu, J.\ Cayao, and M.\ Thakurathi for valuable discussions.

\appendix

\section{Energy gap structure and the emergence of nodal structures in PUM-SCs}

We demonstrate the energy gap structure in the bulk and the condition for the emergence of nodal structures of $p$-wave unconventional magnet/$s$-wave SC hybrid systems (PUM-SC) in Appendix A.
First of all, we calculate the eigenvalues of the Bogoliubov-de Gennes Hamiltonian.
We mentioned that the separated Hamiltonian in the normal state can be obtained as Eq.\ (\ref{PUM_eq_2}) in Section II.
Likewise, the Bogoliubov-de Gennes Hamiltonian in Eq.\ (\ref{BdG_eq}) can also be separated into two subspaces $\eta=\pm 1$: 
\begin{align}
    \hat{H}^{\eta}_\mathrm{BdG}(\bm{k})&=\varepsilon(\bm{k})\hat{s}_{0}\otimes\hat{\tau}_{3}+[t_{x}\sin{k_x}+t_{y}\sin{k_y}]\hat{s}_{3}\otimes\hat{\tau}_{0}\notag\\
    &+\eta J\hat{s}_{1}\otimes\hat{\tau}_{3}+\Delta i\hat{s}_{2}\otimes i\hat{\tau}_{2},
\end{align}%
with the Pauli matrices in Nambu space $\hat{\tau}_{0,1,2,3}$.
By the diagonalization of the above Bogoliubov-de Gennes Hamiltonian, for each sector $\eta=\pm1$, we obtain the eigenvalues at $t_{y}=0$ ($p_{x}$-wave UM-SCs):
\begin{widetext}
    \begin{align}
        E^{s}_{\pm}(\bm{k})=\pm\sqrt{\varepsilon^{2}(\bm{k})+J^{2}+t^2_{x}\sin^{2}{k_x}+\Delta^{2}+2s\sqrt{\varepsilon^{2}(\bm{k})J^{2}+\varepsilon^{2}(\bm{k})t^{2}_{x}\sin^{2}{k_x}+J^{2}\Delta^{2}}},\label{BdG_Ek_x}
    \end{align}%
    with $s=\pm1$ and
    \begin{align}
        \varepsilon(\bm{k})=-\mu-2t\cos{k_x}-2t\cos{k_y},
    \end{align}%
    and at $t_{x}=0$ ($p_{y}$-wave UM-SCs):
    \begin{align}
        E^{s}_{\pm}(\bm{k})=\pm\sqrt{\varepsilon^{2}(\bm{k})+J^{2}+t^2_{y}\sin^{2}{k_y}+\Delta^{2}+2s\sqrt{\varepsilon^{2}(\bm{k})J^{2}+\varepsilon^{2}(\bm{k})t^{2}_{y}\sin^{2}{k_y}+J^{2}\Delta^{2}}}.
    \end{align}%
\end{widetext}%

\begin{figure}[t!]
    \centering
    \includegraphics[width=8.5cm]{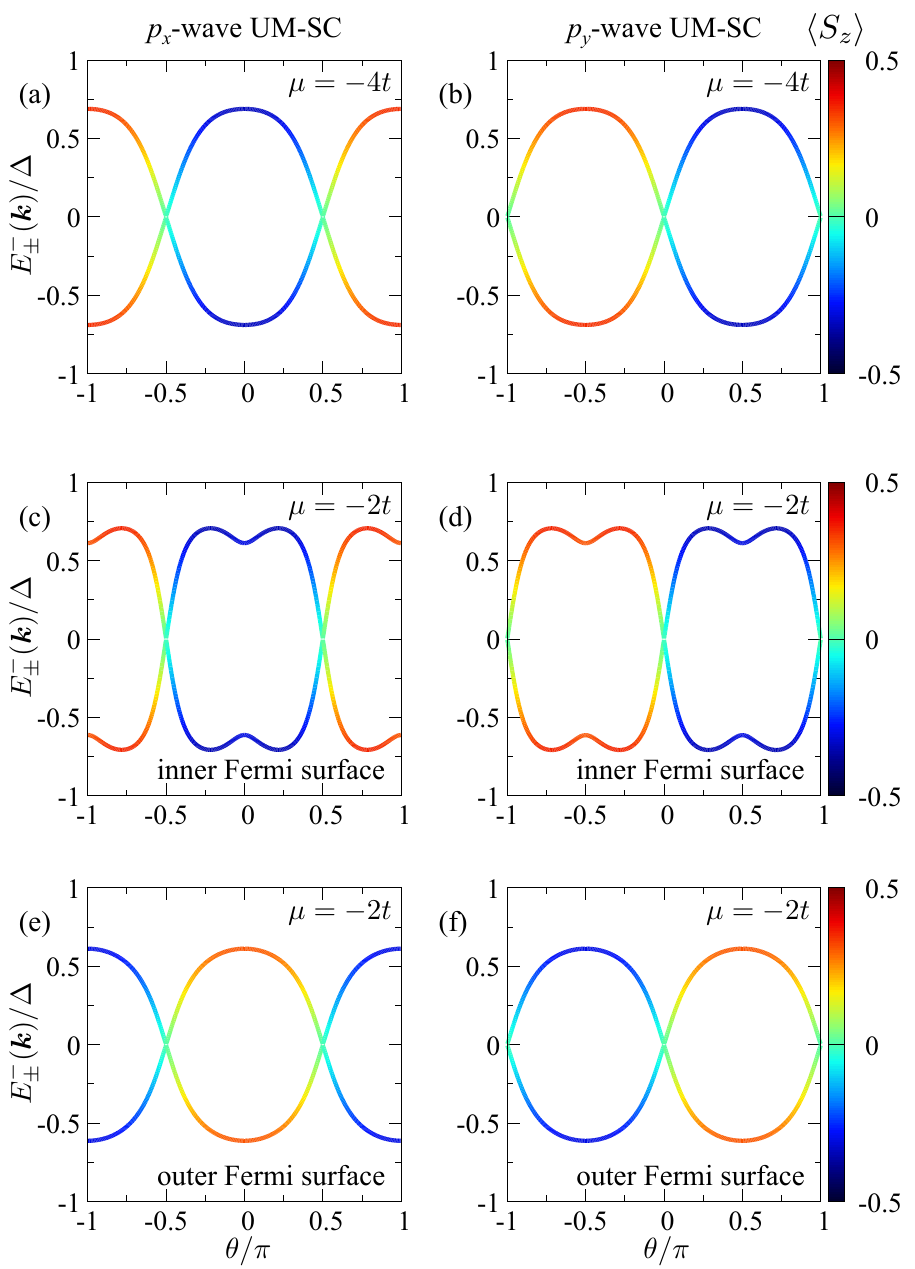}
    \caption{The lowest quasiparticle energy dispersion $E^{-}_{\pm}(\bm{k})$ on the Fermi surface, with the expectation values of $\hat{s}_{z}$ ($\langle S_{z}\rangle$ in the nomral state at (a)(c)(e) $(t_{x},t_{y})=(t,0)$ and (b)(d)(f) $(t_{x},t_{y})=(0,t)$.
    We select the chemical potential as (a)(b) $\mu=-4t$ and (c)-(f) $\mu=-2t$.
    $\theta$ means the angle in the $(k_x,k_y)$ coordinate.
    Panels (c) and (d) [(e) and (f)] are for the inner (outer) Fermi surface shown in Fig.~\ref{fig:1} (d) and Fig.~\ref{fig:1} (f), respectively.
    Exact nodal points do not appear on the Fermi surface; that is, point nodes misalign from the Fermi surface.
    Parameters: $J=t$ and $\Delta=0.01t$.}
    \label{figa:1}
\end{figure}

Based on these equations, we can find that quasiparticle energy dispersions are doubly degenerate for sectors $\eta=\pm 1$.
For the $p_{x}$-wave UM-SC $(t_{x},t_{y})=(t,0)$, we plot the lowest quasiparticle energy dispersions $E^{-}_{\pm}(\bm{k})$ on the Fermi surface at $\mu=-4t$ [Fig.~\ref{figa:1} (a)] and $\mu=-2t$ [Figs.~\ref{figa:1} (c)(e)], at $J=t$.
Then we also show the expectation value of $S_{z}=(\hbar/2)\hat{s}_{z}$ ($\langle S_{z}\rangle$) in the normal state.
Point nodes appear at on the $k_{y}$-axis ($\theta=\pm\pi/2$) with $\langle S_{z}\rangle=0$.
As well as for $p_{y}$-wave UM-SC $(t_{x},t_{y})=(0,t)$ and $J=t$, we also obtain point nodes on the $k_x$-axis ($\theta=0,\pi$) and then $\langle S_{z}\rangle$ vanishes [Figs.~\ref{figa:1} (b)(d)(f)].

\begin{figure}[t!]
    \centering
    \includegraphics[width=8.5cm]{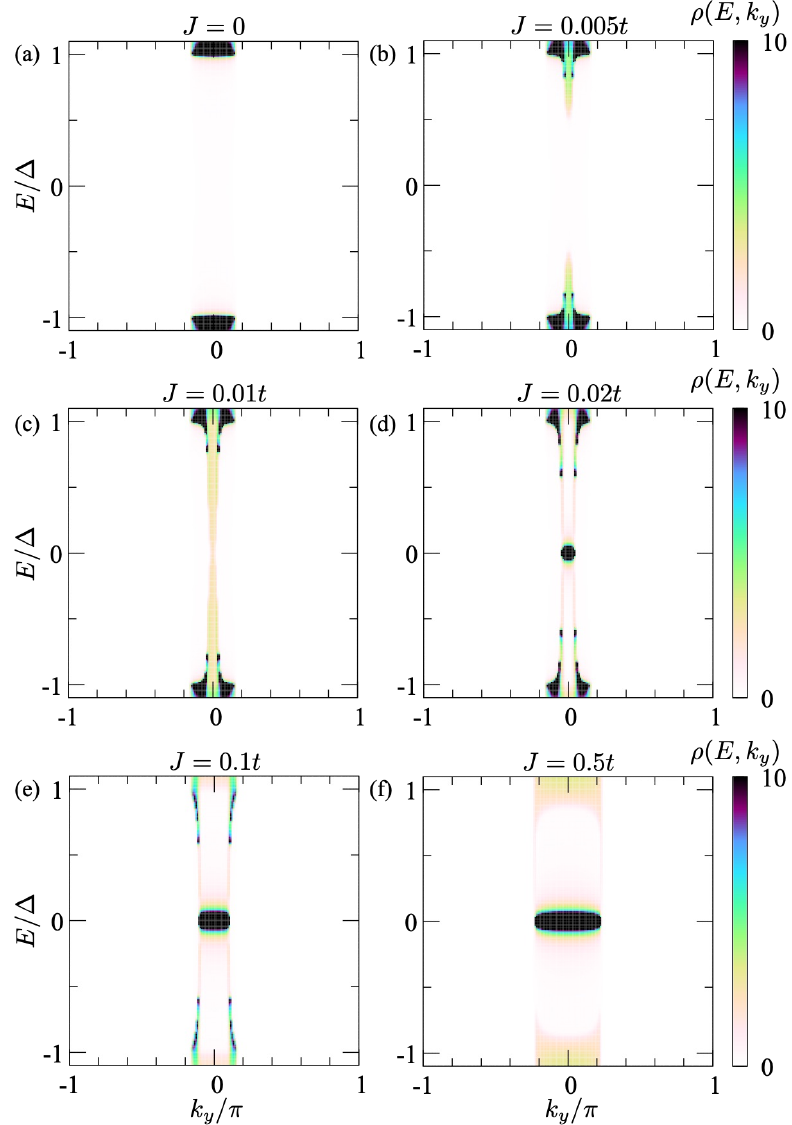}
    \caption{Momentum-resolved surface density of states at the [100] edge $D(E)$ by changing $J$ at $(t_{x},t_{y})=(t,0)$ ($p_x$-wave UM-SC), $\mu=-4t$, and $\Delta=0.01t$.
    We select $J$ as (a) $J=0$, (b) $J=0.005t$, (c) $J=0.01t$, (d) $J=0.02t$, (e) $J=0.1t$, and (f) $J=0.5t$.
    Nodal structures are formed at $J=\Delta$ in panel (c), and zero-energy flat bands emerge for $J>\Delta$ in panel (d)(e)(f).
    Similar behavior is also obtained at $\mu=-2t$.}
    \label{figa:2}
\end{figure}

Next, we confirm the condition of the emergence of nodal structures in the bulk of PUM-SCs at $(t_{x},t_{y})=(t,0)$ ($p_{x}$-wave UM-SC), $\mu=-4t$, and $\Delta=0.01t$.
At $J=0$ [Fig.~\ref{figa:2} (a)], the fully-gapped structure appears. 
As shown in Fig.~\ref{figa:2} (b) at $J=0.005t$, for $J<\Delta$, the energy gap structure is still fully-gapped.
At $J=\Delta=0.01t$, the nodal structure is obtained [Fig.~\ref{figa:2} (c)], and zero-energy flat bands are also realized at $J=0.02t,0.10t,0.50t$ [Fig.~\ref{figa:2} (d)(e)(f)], for $J>\Delta$.
Then, in Eq.\ (\ref{BdG_Ek_x}), we can find the condition of the realization of the gap closing: $J=\Delta$ for $E^{-}_{\pm}(\bm{k})=0$ and $\varepsilon(\bm{k})=0$, at $k_x=0$.
Thus, for $J>\Delta$ with nonzero $t_{x,y}$, nodal structures with zero-energy flat bands can be exhibited in the bulk of PUM-SCs.


\begin{figure}[t!]
    \centering
    \includegraphics[width=8.5cm]{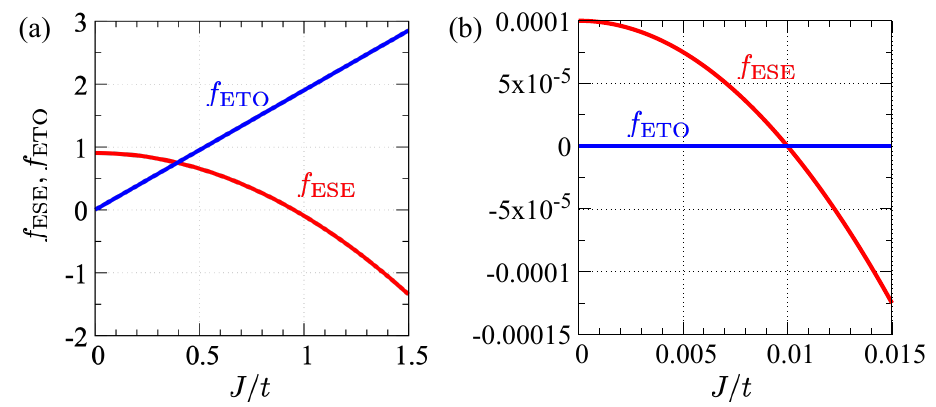}
    \caption{The value of the numerators $f_\mathrm{ESE}=t^2\sin^{2}{k_\mathrm{F}}-J^2+\Delta^{2}$ and $f_\mathrm{ETO}=2Jt\sin{k_\mathrm{F}}$ at (a) $k_{\mathrm{F}x}=0.4\pi$ and (b) $k_{\mathrm{F}x}=0$, as a function of $J$. 
    We select the parameters as $t=1$ and $\Delta=0.01t$.}
    \label{reply_ff}
\end{figure}%

\section{Analytical solution of the anomalous Green's functions of PUM-SCs in the bulk}

In Appendix B, we discuss the analytical solution of the anomalous Green's functions (pair amplitude) of PUM-SCs in the bulk.
Green's function in the particle-hole Nambu space is calculated as:
\begin{align}
    \tilde{G}_{\eta}(\bm{k},i\omega_{n})&=[i\omega_{n}-\hat{H}_{\eta}(\bm{k})]^{-1}\notag\\
    &=\begin{pmatrix}
        \hat{G}(\bm{k},i\omega_{n}) & \hat{F}(\bm{k},i\omega_{n})\\
        \bar{F}(\bm{k},i\omega_{n}) & \bar{G}(\bm{k},i\omega_{n})
    \end{pmatrix},
\end{align}%
with the Fermionic Matsubara frequency $\omega_{n}=(2n+1)\pi k_\mathrm{B}T$ and the temperature $T$.
When we focus on $(t_{x},t_{y})=(t,0)$ ($p_x$-wave UM), the nonzero pair amplitude is:
\begin{align}
    F^{\uparrow\downarrow}_\mathrm{ESE}(\bm{k},i\omega_{n})&=-F^{\downarrow\uparrow}_\mathrm{ESE}(\bm{k},i\omega_{n})\label{ESE_x}\\
    &=\frac{\Delta[(\varepsilon(\bm{k})-t_{x}\sin{k_x})^2-J^{2}+\omega^{2}_{n}+\Delta^{2}]}{D_{x}},\notag
\end{align}%
for ESE,
\begin{align}
    F^{\uparrow\uparrow}_\mathrm{ETO}=F^{\downarrow\downarrow}_\mathrm{ETO}=\frac{2\Delta \eta Jt_x\sin{k_x}}{D_{x}},\label{ETO_x}
\end{align}%
for ETO, 
\begin{align}
    F^{\uparrow\downarrow}_\mathrm{OSO}(\bm{k},i\omega_{n})&=F^{\downarrow\uparrow}_\mathrm{OSO}(\bm{k},i\omega_{n})=0,
\end{align}%
for OSO~\cite{BalatskyPRB1992,LinderRMP2019}, and
\begin{align}
    F^{\uparrow\uparrow}_\mathrm{OTE}=F^{\downarrow\downarrow}_\mathrm{OTE}=\frac{2i\Delta \eta J \omega_{n}}{D_{x}},\label{OTE_x}
\end{align}%
for OTE pairings~\cite{berezinskii1974,KirkpatrickPRL1991,BergeretPRB2005,LinderRMP2019}, with
\begin{align}
    D_{x}&=t^{4}_{x}\sin^{4}{k_x}+\Delta^{4}+\varepsilon^{4}(\bm{k})+J^{2}\notag\\
    &-2\varepsilon^{2}(\bm{k})J^{2}+2\varepsilon^{2}(\bm{k})\omega^{2}_{n}+2J^{2}\omega^{2}_{n}+\omega^{4}_{n}\notag\\
    &+2\Delta^{2}[\varepsilon^{2}(\bm{k})-J^{2}+\omega^{2}_{n}]\notag\\
    &+2t^{2}_{x}\sin^{2}{k_x}[\Delta^{2}-\varepsilon^{2}(\bm{k})+J^{2}+\omega^{2}_{n}],
\end{align}%
As well as at $(t_{x},t_{y})=(0,t)$ ($p_y$-wave UM), we obtain
\begin{align}
    F^{\uparrow\downarrow}_\mathrm{ESE}(\bm{k},i\omega_{n})&=-F^{\downarrow\uparrow}_\mathrm{ESE}(\bm{k},i\omega_{n})\\
    &=\frac{\Delta[(\varepsilon(\bm{k})-t_{y}\sin{k_y})^2-J^{2}+\omega^{2}_{n}+\Delta^{2}]}{D_{y}},\notag
\end{align}%
for ESE,
\begin{align}
    F^{\uparrow\uparrow}_\mathrm{ETO}(\bm{k},i\omega_{n})=F^{\downarrow\downarrow}_\mathrm{ETO}(\bm{k},i\omega_{n})=\frac{2\Delta \eta Jt_y\sin{k_y}}{D_{y}},
\end{align}%
for ETO, 
\begin{align}
    F^{\uparrow\downarrow}_\mathrm{OSO}(\bm{k},i\omega_{n})&=F^{\downarrow\uparrow}_\mathrm{OSO}(\bm{k},i\omega_{n})=0,
\end{align}%
for OSO, and
\begin{align}
    F^{\uparrow\uparrow}_\mathrm{OTE}(\bm{k},i\omega_{n})=F^{\downarrow\downarrow}_\mathrm{OTE}(\bm{k},i\omega_{n})=\frac{2i\Delta \eta J \omega_{n}}{D_{y}}, \label{OTE_y}
\end{align}%
for OTE pairings, with
\begin{align}
    D_{y}&=t^{4}_{y}\sin^{4}{k_y}+\Delta^{4}+\varepsilon^{4}(\bm{k})+J^{2}\notag\\
    &-2\varepsilon^{2}(\bm{k})J^{2}+2\varepsilon^{2}(\bm{k})\omega^{2}_{n}+2J^{2}\omega^{2}_{n}+\omega^{4}_{n}\notag\\
    &+2\Delta^{2}[\varepsilon^{2}(\bm{k})-J^{2}+\omega^{2}_{n}]\notag\\
    &+2t^{2}_{y}\sin^{2}{k_y}[\Delta^{2}-\varepsilon^{2}(\bm{k})+J^{2}+\omega^{2}_{n}].
\end{align}%
Eqs.\ (\ref{OTE_x}) and (\ref{OTE_y}) indicate that OTE pairing is induced by nonzero $J$~\cite{BergeretPRB2005}.
In addition, to compare with the ESE and ETO pairings, because the denominator is the same in Eqs.\ (\ref{ESE_x}) and (\ref{ETO_x}), we can focus on the numerator to evaluate the sale of the pair amplitude.
In Eq.\ (\ref{ESE_x}) (ESE pairing), at the Fermi level at $(k_{x},k_{y})=(k_{\mathrm{F}x},k_{\mathrm{F}y})$, the scale of $\varepsilon(\bm{k})$ becomes 0, that is, $\varepsilon(\bm{k})\sim 0$.
Then, for $t_{x}=t$, if we do not see the Matsubara frequency term $\omega_n$, the order of the numerator in Eq.\ (\ref{ESE_x}) can be approximated as $f_\mathrm{ESE}=\Delta(t^2\sin^{2}{k_\mathrm{F}}-J^2+\Delta^{2})$.
Likewise, in Eq.\ (\ref{ETO_x}) (ETO pairing), for $t_{x}=t$, we can obtain the scale $f_\mathrm{ETO}=2\Delta Jt\sin{k_\mathrm{F}}$.
Indeed, the ETO state with equal spins can be dominant in the bulk, not the spin-singlet $s$-wave.
We plot $f_\mathrm{ESE}$ and $f_\mathrm{ETO}$ as a function of $J$ at Fig.~\ref{reply_ff} (a) $k_{\mathrm{F}x}=0.4\pi$ and Fig.~\ref{reply_ff} (b) $k_{\mathrm{F}x}=0.0$.
We can find $\Delta(t^2\sin^{2}{k_\mathrm{F}}-J^2+\Delta^{2})<2\Delta Jt\sin{k_\mathrm{F}}$ at $k_{\mathrm{F}x}=0.4\pi$ for $J>0.4t$ in Fig.~\ref{reply_ff} (a), and then the pair amplitude for ETO pairing becomes larger than ESE pairing.
At $k_{\mathrm{F}x}=0.0$, in Fig.~\ref{reply_ff} (b), $f_\mathrm{ESE}$ also becomes zero at $J=\Delta$ with $\Delta=0.01t$, and we obtain $f_\mathrm{ETO}=0$.
This corresponds to the formation of the nodal structure at $k_y=0$ shown in Fig.~\ref{figa:2} (c).
Although we find that nodal structures with zero-energy flat bands appear for $J>\Delta$ discussed in Appendix A, ETO pairing is mainly realized in the bulk by both $J>0.4t$ at $(t_{x},t_{y})=(t,0)$.

\begin{figure}[t!]
    \centering
    \includegraphics[width=8.5cm]{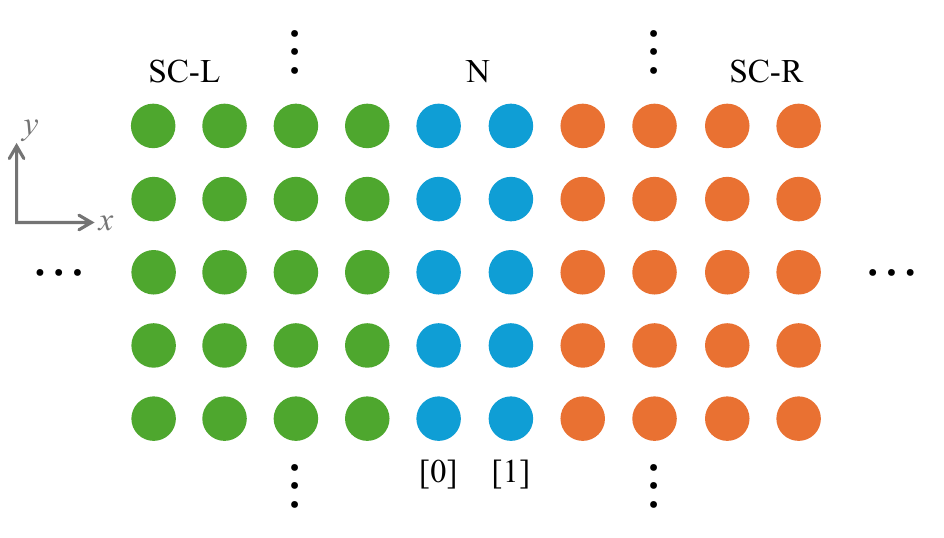}
    \caption{Schematic illustration of Josephson junctions in the lattice model.
    SC-L, N, and SC-R indicate the left-side superconductor (SC), the normal metal, and the right-side SC, respectively.
    [0] and [1] mean the lattice indices along the $x$-direction: $j_x=0$ and $1$, respectively.}
    \label{figa:3}
\end{figure}%

\section{Recursive Green's function method in Josephson junctions}

Here, we provide how we calculate the Josephson current in the present study.
We deal with the tight-binding model with the periodic boundary condition along the $y$-direction and semi-infinite SCs on both sides of the junctions shown in Fig.~\ref{figa:3}.
Then we consider the pair potential as $\hat{\Delta}_\mathrm{L}=\hat{\Delta}$ on the left and $\hat{\Delta}_\mathrm{R}=\hat{\Delta}e^{-i\varphi}$ on the right-side SCs with the phase difference $\varphi$.

\begin{figure}[t!]
    \centering
    \includegraphics[width=8.6cm]{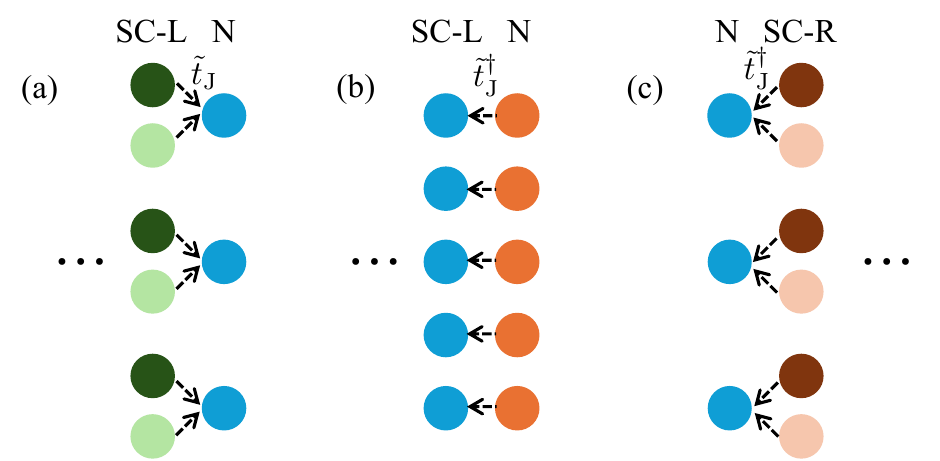}
    \caption{Schematic illustrations of tunneling processes at the interface for (a) $p$-wave unconventional magnet (PUM)-$s$-wave SC hybrid system/normal metal, (b) normal metal/conventional $s$-wave SC, and (c) normal metal/PUM-SC interfaces.
    SC-L, N, and SC-R indicate the left-side SC, the normal metal, and the right-side SC, respectively.
    In (a) and (c), dark and light colors correspond to the sector degrees of freedom in PUM.
    The black arrows mean the recursive directions for the tunneling Hamiltonian $\tilde{t}_\mathrm{J}$. }
    \label{figa:4}
\end{figure}%

First of all, for $z=i\omega_{n}$, we obtain the semi-infinite Green's function in the left $\tilde{G}_\mathrm{L}(k_y,i\omega_{n})$ and right-side SCs $\tilde{G}_\mathrm{R}(k_y,i\omega_{n})$, see also in Ref.\ \cite{Umerski97,TakagiPRB2020}.
By using the recursive Green's function method~\cite{Umerski97,fukaya2024}, we calculate the surface Green's function at $j_x=0,1$:
\begin{align}
    &\tilde{G}^{(0)}_\mathrm{L}(k_y,i\omega_{n})\notag\\
    &=[i\omega_{n}-\tilde{u}_\mathrm{N}(k_y)-\tilde{t}^{\dagger}_\mathrm{J}\tilde{G}_\mathrm{L}(k_y,i\omega_{n})\tilde{t}_\mathrm{J}]^{-1},\\
    &\tilde{G}^{(1)}_\mathrm{R}(k_y,i\omega_{n})\notag\\
    &=[i\omega_{n}-\tilde{u}_\mathrm{N}(k_y)-\tilde{t}_\mathrm{J}\tilde{G}_\mathrm{R}(k_y,i\omega_{n})\tilde{t}^{\dagger}_\mathrm{J}]^{-1},
\end{align}%
with the local term in the normal metal $\tilde{u}_\mathrm{N}(k_y)$.
Here, $\tilde{t}_\mathrm{J}$ denotes the tunneling Hamiltonian:
\begin{align}
    \tilde{t}_\mathrm{J}&=t_\mathrm{int}
    \begin{pmatrix}
        -t & 0 \\
        -t & 0 \\
         0 & -t \\
         0 & -t
    \end{pmatrix}\otimes\hat{\tau}_{3},
\end{align}%
for the PUM-SC/normal metal interface [Fig.~\ref{figa:4} (a)], 
\begin{align}
    \tilde{t}_\mathrm{J}&=t_\mathrm{int}
    \begin{pmatrix}
        -t & 0 \\
        0  & -t 
    \end{pmatrix}\otimes\hat{\tau}_{3},
\end{align}%
for the normal metal/conventional $s$-wave SC interface [Fig.~\ref{figa:4} (b)], and
\begin{align}
    \tilde{t}_\mathrm{J}&=t_\mathrm{int}
    \begin{pmatrix}
        -t & -t &  0 & 0\\
         0 &  0 & -t &-t
    \end{pmatrix}\otimes\hat{\tau}_{3},
\end{align}%
for the normal metal/PUM-SC interface [Fig.~\ref{figa:4} (c)], where $t_\mathrm{int}$ is the transparency amplitude at the interface.
In junctions at $j_{x}=0,1$, we obtain the local and nonlocal Green's functions:
\begin{align}
    &\tilde{G}_{0,0}(k_y,i\omega_{n})\notag\\
    &=[\{\tilde{G}^{(0)}_\mathrm{L}(k_y,i\omega_{n})\}^{-1}-\tilde{t}_\mathrm{N}\tilde{G}^{(1)}_\mathrm{R}(k_y,i\omega_{n})\tilde{t}^{\dagger}_\mathrm{N}],\\
    &\tilde{G}_{1,1}(k_y,i\omega_{n})\notag\\
    &=[\{\tilde{G}^{(1)}_\mathrm{R}(k_y,i\omega_{n})\}^{-1}-\tilde{t}^{\dagger}_\mathrm{N}\tilde{G}^{(0)}_\mathrm{L}(k_y,i\omega_{n})\tilde{t}_\mathrm{N}],
\end{align}%
\begin{align}
    \tilde{G}_{0,1}(k_y,i\omega_{n})&=\tilde{G}^{(0)}_\mathrm{L}(k_y,i\omega_{n})\tilde{t}_\mathrm{N}\tilde{G}_{1,1}(k_y,i\omega_{n}),\\
    \tilde{G}_{1,0}(k_y,i\omega_{n})&=\tilde{G}^{(1)}_\mathrm{R}(k_y,i\omega_{n})\tilde{t}^{\dagger}_\mathrm{N}\tilde{G}_{0,0}(k_y,i\omega_{n}),
\end{align}%
where $\tilde{t}_\mathrm{N}$ is the nearest-neighbor hopping term in the normal metal:
\begin{align}
    \tilde{t}_\mathrm{N}&=
    \begin{pmatrix}
        -t & 0 \\
        0 & -t
    \end{pmatrix}\otimes\hat{\tau}_{3}.
\end{align}
Thus, the Josephson current can be calculated by the nonlocal Green's functions in Eq.\ (\ref{JC}).

\begin{figure}[t!]
    \centering
    \includegraphics[width=8.5cm]{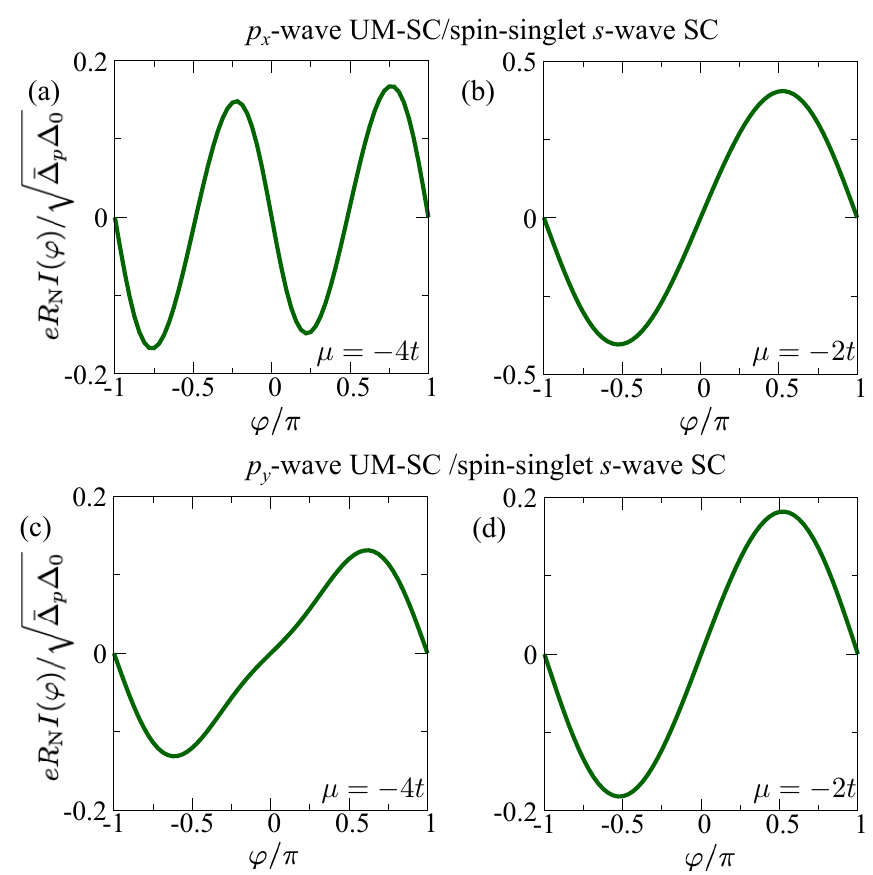}
    \caption{Current phase relation in PUM-SC/spin-singlet $s$-wave SC Josephson junctions at (a)(b) $(t_{x},t_{y})=(t,0)$ and (c)(d) $(t_{x},t_{y})=(0,t)$.
    We select the chemical potential as (a)(c) $\mu=-4t$ and (b)(d) $\mu=-2t$.
    We choose the critical temperature in PUM-SC as $p=0.1$.
    Parameters: $\mu_\mathrm{N}=-3.5t$, $\mu_{s}=-1.5t$, $J=t$, $T_\mathrm{c}=0.01t$, $T=0.025T_{\mathrm{c}}$, and $t_\mathrm{int}=1$.}
    \label{figa:5}
\end{figure}%
\begin{figure}[t!]
    \centering
    \includegraphics[width=8.5cm]{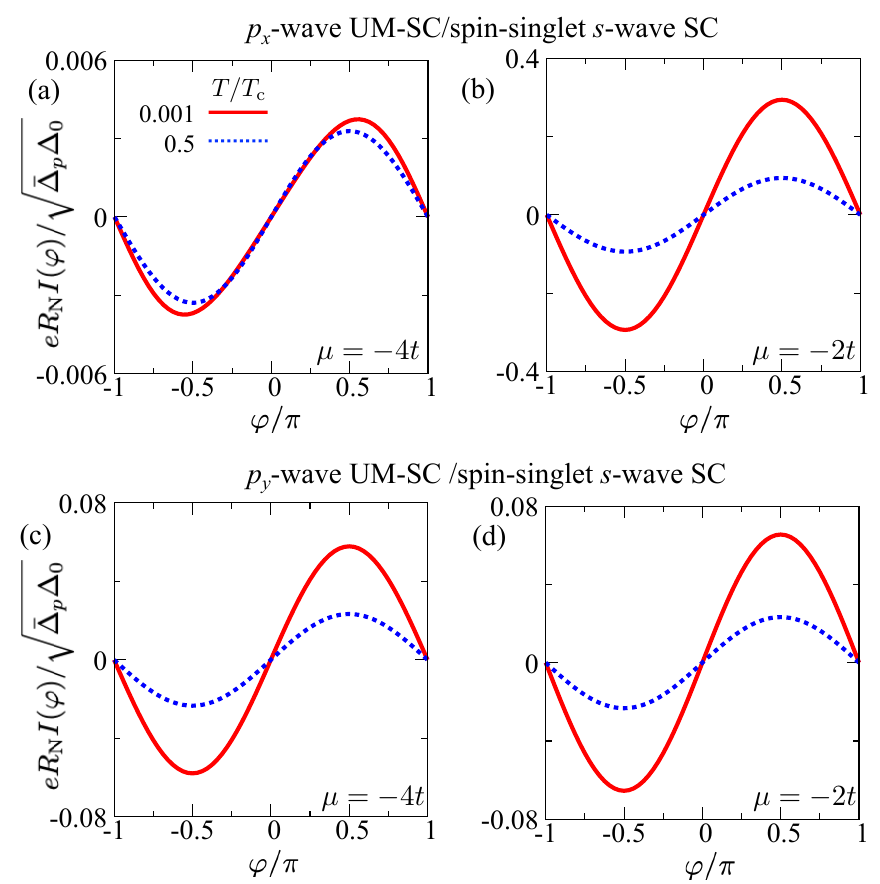}
    \caption{Current phase relation in PUM-SC/spin-singlet $s$-wave SC Josephson junctions for $T=0.001T_{\mathrm{c}}$ (red-solid line) and $T=0.5T_{\mathrm{c}}$ (blue-dotted line) at (a)(b) $(t_{x},t_{y})=(t,0)$ and (c)(d) $(t_{x},t_{y})=(0,t)$.
    We select the chemical potential as (a)(c) $\mu=-4t$ and (b)(d) $\mu=-2t$.
    We choose the critical temperature in PUM-SC as $p=0.1$.
    Parameters: $\mu_\mathrm{N}=-3.5t$, $\mu_{s}=-1.5t$, $J=t$, $T_\mathrm{c}=0.01t$, $T=0.025T_{\mathrm{c}}$, and $t_\mathrm{int}=0.1$.}
    \label{figa:6}
\end{figure}%

\section{Josephson current: effect of the smaller magnitude of the pair potential in PUM-SC}

Because the effective amplitude of the pair potential in PUM-SC is different from that in spin-singlet $s$-wave SCs, in Appendix D, we provide the current phase relation when we choose the smaller magnitude of the pair potential in PUM-SC/spin-singlet $s$-wave SC Josephson junctions.
To analyze the influence of the amplitude of the pair potential in PUM-SC, we demonstrate the Josephson current by choosing $\bar\Delta_{p}=p\Delta_{0}$ with $p=0.1$.
We plot the current phase relation at the high-transparency in PUM-SC/$s$-wave SC junctions in Fig.~\ref{figa:5}.
We calculate the current phase relation at $p=0.1$, $T=0.025T_\mathrm{c}$, and $T_\mathrm{c}=0.01t$ in Fig.~\ref{figa:5}, for each chemical potential at $\mu=-4t$ [Fig.~\ref{figa:5} (a) and Fig.~\ref{figa:5} (c)] and $\mu=-2t$ [Fig.~\ref{figa:5} (b) and Fig.~\ref{figa:5} (d)].
Although the amplitude of the maximum Josephson current becomes small, we obtain similar results qualitatively at $p=0.1$ [Fig.~\ref{figa:5}], comparing the original results in Figs.~\ref{fig:6} (b)-(e).
At the low transparency, we show the current phase relation in Fig.~\ref{figa:6}. 
The non-sinusoidal feature disappears at $T=0.001T_\mathrm{c}$ and $(t_x,t_y,\mu)=(t,0,-4t)$ in Fig.~\ref{figa:6} (a), see also Fig.~\ref{fig:9} (a).
We get the similar results qualitatively at $p=0.1$ [Figs.~\ref{figa:6} (b-d)], as compared with at $p=1$ [Figs.~\ref{fig:9} (b-d)].
Hence, the current phase relations, as well as the temperature dependence, can be influenced by the choice of $\bar{\Delta}_{p}$.

\bibliography{biblio}


\end{document}